\documentclass[aps, reprint, groupedaddress,  longbibliography, superscriptaddress, amsmath, amssymb]{revtex4-1} 
\usepackage[colorlinks,allcolors=blue]{hyperref}
\usepackage{graphicx}
\usepackage{epstopdf}
\usepackage{dcolumn}
\usepackage{bm}
\usepackage{braket}
\usepackage{tikz}
\usetikzlibrary{shapes, arrows, shadows}
\usepackage{braket}

\newcommand{\beginsupplement}{%
    \setcounter{table}{0}
    \renewcommand{\thetable}{\arabic{table}}%
    \setcounter{figure}{0}
    \renewcommand{\thefigure}{\arabic{figure}}%
    \setcounter{equation}{0}
    \renewcommand{\theequation}{\arabic{equation}}%
}

\newcommand{\TITLE}{\textbf{Ultrafast charge ordering by self-amplified exciton-phonon dynamics in TiSe$_2$}}

\begin{document}
\title{\TITLE}

\author{Chao Lian}
\affiliation{Beijing National Laboratory for Condensed Matter
	Physics and Institute of Physics, Chinese Academy of Sciences,
	Beijing, 100190, P. R. China}

\author{Sheng-Jie Zhang}
\author{Shi-Qi Hu}
\author{Meng-Xue Guan}
\affiliation{Beijing National Laboratory for Condensed Matter
	Physics and Institute of Physics, Chinese Academy of Sciences,
	Beijing, 100190, P. R. China}

\author{Sheng Meng}
\affiliation{Beijing National Laboratory for Condensed Matter
Physics and Institute of Physics, Chinese Academy of Sciences,
Beijing, 100190, P. R. China}
\affiliation{School of Physical Sciences, University of Chinese Academy of Sciences,
	Beijing, 100190, P. R. China}
\affiliation{Songshan Lake Materials Laboratory, Dongguan, Guangdong 523808, P. R. China}

\date{\today}

\begin{abstract}
The origin of charge density waves (CDW) in TiSe$_2$ has long been debated, mainly due to the difficulties in identifying the timescales of how and when the excitonic pairing and electron-phonon coupling (EPC) come into play. Without a proper time resolution and microscopic mechanism, one has to assume simultaneous appearance of CDW and periodic lattice distortions (PLD). Here, we accomplish a complete separation of exciton and PLD dynamics and unravel their interplay in the ultrafast time domain in our real-time time-dependent density functional theory simulations. We find that laser pulses knock off the exciton order and induce a homogeneous bonding-antibonding transition in the initial 20 fs, then the weakened electronic order triggers ionic movements antiparallel to the original PLD. The EPC comes into play after the initial 20~fs, and the two processes mutually amplify each other leading to a complete inversion of CDW ordering. The self-amplified dynamics reproduces the evolution of band structures in excellent agreement with ultrafast photoemission experiment. Hence we resolve the key processes in the initial dynamics of CDW that help elucidate the mechanism underlying the long debated problem.
\end{abstract}

\maketitle

Charge density wave (CDW) in 1T-TiSe$_2$ has been one of the persistent eye-drawing topics over decades. It is not only an excellent playground to study the interplay between CDW and superconductivity~\cite{Morosan2006, Li2007,  Kusmartseva2009, Yao2018, Wei2017b, Kogar2017a, Yan2017, Medvecka2016, Das2015, Luna2015, Ganesh2014, Joe2014, Kacmarcik2013, Husanikova2013, Iavarone2012, Jeong2010, Zaberchik2010, Giang2010, Morosan2010, Hillier2010, Kusmartseva2009, Barath2008, Li2007b, Li2007, Morosan2006, Cui2006}, but also an evidenced excitonic insulator~\cite{Kohn1967, Jerome1967, Halperin1968, Rossnagel2002, Cercellier2007}. It was heavily debated which mechanism -- the electron-phonon coupling (EPC)~\cite{Hughes1977, Wakabayashi1978, Gaby1981, Motizuki1981, Lopez-Castillo1987, Holt2001, Bussmann-Holder2009, VanWezel2010a, Rossnagel2010, Calandra2011, Zhu2012, Zenker2013} or the excitonic pairing~\cite{Kidd2002, Monney2012a, VanWezel2010, Stoffel1982,  Monney2015, Cazzaniga2012, Monney2010, Anderson1985a, Zenker2013, Peng2015b, Koley2014, Novello2017, Monney2012, Monney2010a, Monney2011, Monney2009, Sugawara2016, Monney2012a, Hildebrand2016a, Watanabe2015, Pillo2000, May2011,Chen2018c} -- is the major driving force for the formation of CDW in TiSe$_2$. 
These two mechanisms disagree on the role of periodic lattice distortions (PLD) in CDW: PLD is essential in forming the CDW according to the EPC mechanism, while it is only a passive consequence of the CDW if the excitonic pairing dominates.

Isolating the PLD from the CDW can solve the debate, but it is not achievable in the ground state. Inspiringly, the ultrafast measurements can distinguish PLD and CDW based on their different time scales~\cite{Sundaram2002}. However, conclusions from previous ultrafast measurements are highly controversial. Utilizing time-resolved (tr) and angle-resolved photoemission spectroscopy (ARPES) measurements, Rohwer~\textit{et al.} observed very fast ($<30$~fs) collapses of the long-range order~\cite{Rohwer2011a}, which was interpreted by Mathias~\textit{et al.} as the signals of the mutually-amplified carrier multiplication and gap quenching~\cite{Mathias2016}. However, this mechanism completely neglects the possible involvements of ionic dynamics and the effects of geometry relaxation. M\"{o}hr-Vorobeva \textit{et al.} observed the nonthermal melting within 250~fs in the ultrafast X-ray measurements and suggested that excitonic pairing generates the CDW~\cite{Mohr-Vorobeva2011}. Hellmann~\textit{et al.} also supported the excitonic mechanism~\cite{Hellmann2012}. They observed that the $100$~fs change in the tr-ARPES signals is comparable to the buildup time of the electron-hole screening in the exciton formation. On the other hand, Porer~\textit{et al.} indicated that excitonic pairing was not the sole driving force of CDW~\cite{Porer2014}. They separately studied the electronic and structural orders via monitoring the characteristic peaks of the two in the transient energy loss spectra. They found that the PLD can persist with the quenched excitonic order. Despite extensive efforts in the past, the conclusions in all these previous studies were mainly derived from indirect mappings between the spectra and the CDW/PLD orders. Notwithstanding the cutting-edge techniques used in these studies, the time-resolved spectra can only provide the averaged response of the material without atomic resolution. With recent significant progresses in real-time (rt) time dependent (TD) density functional theory (DFT) algorithms and computing power~\cite{Runge1984, Bertsch2000, Yabana2006, Otobe2008, Otobe2009, Otobe2016b, Yabana2012, Shinohara2010, Shinohara2010a, Shinohara2012, Sato2015, Sato2015a, Ren2013, Wang2015a}, it become now possible to perform  ultrafast quantum dynamics simulations fully from first principles, to provide a unified atomic-level picture of ultrafast CDW dynamics. 

\begin{figure}
	\centering
	\includegraphics[width=1.0\linewidth]{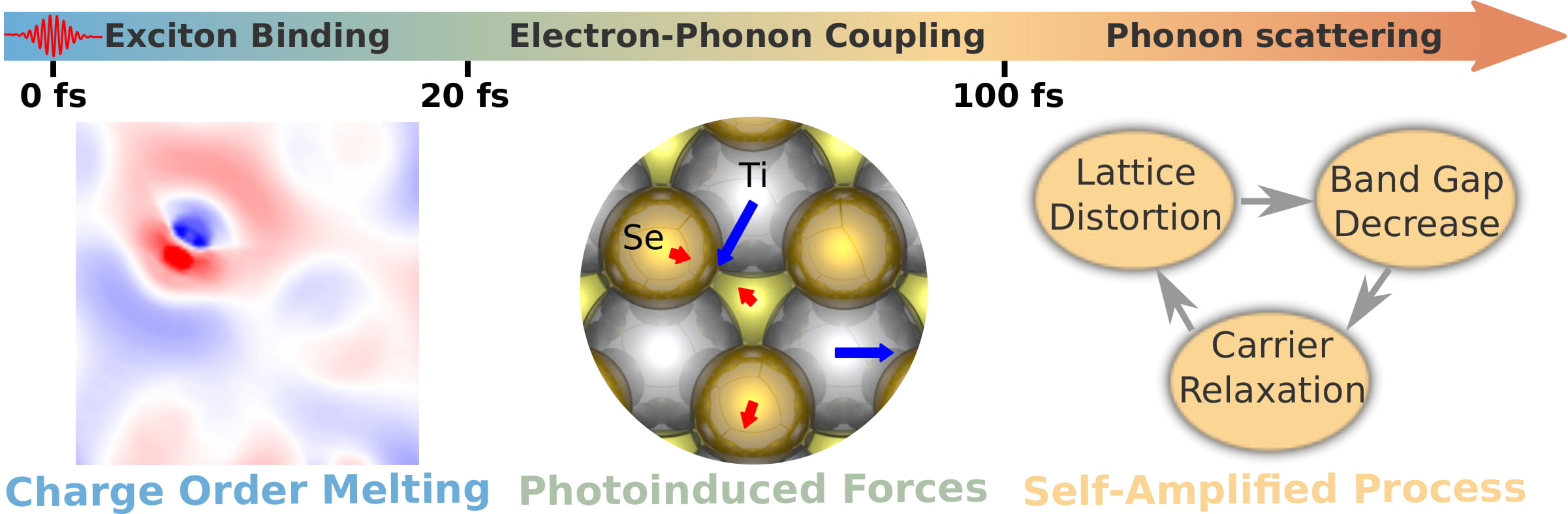}
	\caption{Schematic of atomic processes in photoexcited 1T-TiSe$_2$. The laser pulse melts charge order within 20~fs, producing the forces that trigger the ionic movements. The self-amplified dynamics is assisted by electron-phonon couplings after initial excitation.}
	\label{fig:timescale}
\end{figure}

Here we take advantages of recently developed rt-TDDFT methods for \textit{ab initio} simulations of ultrafast dynamics in complex materials such as TiSe$_2$. We demonstrate that laser pulses knock down the CDW order without disturbing PLD through inducing a homogeneous bonding-antibonding electronic transition. The reduced CDW order then triggers the ionic movements exactly antiparallel to the original PLD, but cannot solely drives the observed inversion in CDW/PLD. Instead, assisted by electron-phonon couplings, a self-amplification mechanism between electron dynamics and lattice distortion emerges after the initial excitation, reproducing well experimental features observed in tr-ARPES measurements. We propose that both exciton pairing and electron-phonon couplings contribute to the CDW formation, albeit in a different timescale (see Fig.~\ref{fig:timescale}):  CDW is predominantly initiated by exciton binding ($<20$~fs) and subsequently enhanced by EPC ($>20$~fs). These insights hint for a complete  microscopic understanding on the nature of charge ordering in quantum materials. 

\begin{figure}
	\centering	\includegraphics[width=1.0\linewidth]{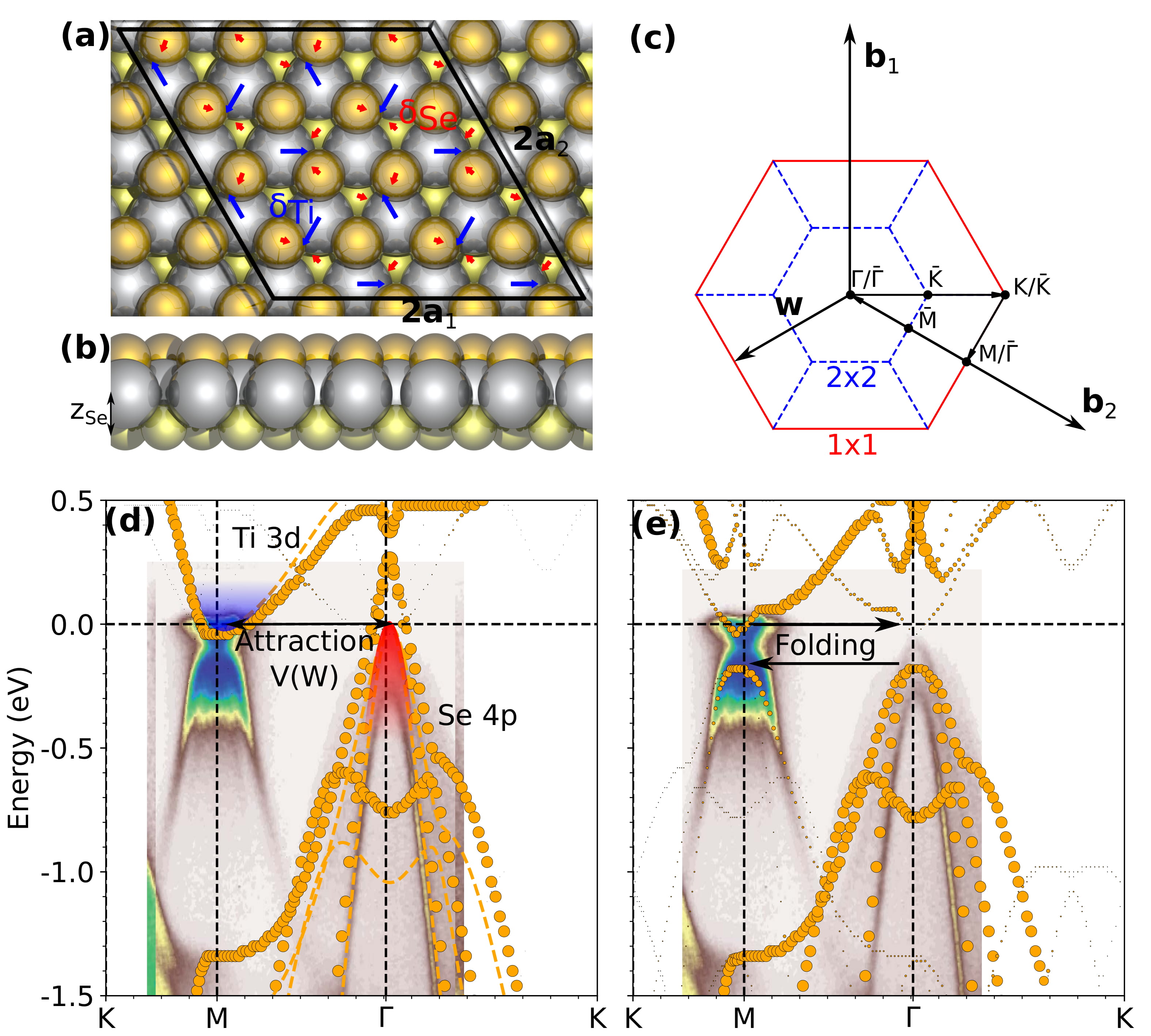}
	\caption{Top view (a) and side view (b) of the atomic structure of 1T-TiSe$_2$. The silver, yellow and orange spheres denote the Ti atoms, the Se atoms on the bottom layer and the Se atoms on the top layer, respectively. The blue and the red arrows denote the PLD displacements \{$\mathbf{d}_i$\} of the Ti and Se atoms, respectively. (c) Brillouin zones (BZ) of TiSe$_2$. The solid red lines and blue dash lines denote the BZ of the $2\times2$ and $1\times1$ cell, respectively. \{$\Gamma$, M, K\} and \{$\bar{\Gamma}$, $\bar{\mathrm{M}}$, $\bar{\mathrm{K}}$\} denote the special $k$ points in $1\times1$ and $2\times2$ BZ, respectively. Effective band structures (EBS) of the (d) normal state with $ U=0$~eV (dots) and $U=3.5$~eV (dashed lines) and (e) CDW state. The plots in shaded areas in (d) and (e) are ARPES spectra reproduced from Ref.~\cite{Rohwer2011a}. The Fermi energies are shifted to the experimental values.
	}
	\label{fig:struct+BZ}
\end{figure}

\vspace{1cm} \noindent \textbf{Results}\\
\textbf{Ground state properties.} We calculate the atomic geometries of the normal phase and the CDW 2$\times$2$\times$2 phase. {The bulk TiSe$_2$ is used with the interlayer separation of 6.69~\AA.}   \label{txt:RefbComl}
The optimized PLD displacements \{$\mathbf{d}_i$\} are shown in Fig.~\ref{fig:struct+BZ}(a), where the displacements $\delta_{Ti} = 0.091$~\AA, $\delta_{Se} = 0.030$~\AA\ and $\delta_{Ti}/\delta_{Se} =3.03:1$. These results excellently reproduce the experimental measurements $\delta_{Ti}=0.085 \pm 0.014$~\AA\ and $\delta_{Ti}/\delta_{Se}\sim3:1$~\cite{DiSalvo1976}.
To directly compare with ARPES measurements, we calculate the effective band structure (EBS) along K-M-$\mathrm{\Gamma}$-K by unfolding energy bands from the $2\times2$ BZ to the $1\times1$ BZ, as shown in Fig.~\ref{fig:struct+BZ}. The calculated EBS agrees well with the experimental spectra. In the normal state, we reproduce the electron and hole pockets at M and $\Gamma$ points, respectively. The CDW opens an indirect gap of 0.18~eV and creates folded valence bands (VB) at M and folded conduction bands (CB) at $\Gamma$. We note that PBE+U with $U_\mathrm{Ti}=3.5$~eV incorrectly indicates that the CDW state is unstable~\cite{Bianco2015}. Thus, we use $U_\mathrm{Ti}=0$~eV for all the dynamics simulations with TDDFT.

Based on the consistency between DFT results and experimental measurements, we briefly discuss the accuracy of DFT in describing the excitonic paring and EPC in TiSe$_2$. {It is well known that the semilocal exchange-correlation (XC) functionals (e.g. PBE) poorly describes the long-range Coulomb screening $2\pi/|\mathbf{k}'-\mathbf{k}| \sim \infty$ in vertical excitations $\mathbf{q}=\mathbf{k}'-\mathbf{k} =0 $~\cite{Sharma2011}. Computationally expensive corrections such as Bethe-Salpeter equation (BSE)~\cite{Sottile2003, Marini2003} can considerably improve the accuracy. The experimentally-observed superlinear feature is absent in our simulations [Supplementary Fig. 2]. However, the long-range attractions produce spatially uniform forces on the ions. Based on the concept of the excitonic insulator~\cite{Kohn1967, Jerome1967, Halperin1968}, the PLD stability is only affected by the inter-valley excitons formed by an attractive interaction $V(\mathbf{w})$ between the electron pocket at the M point and the hole pocket at the $\Gamma$ point, as shown in Fig.~\ref{fig:struct+BZ}(d). Here, $\mathbf{w}=\pm\mathbf{b}_i/2$ and $\mathbf{b}_i$ ($i=1,2$) being the reciprocal lattice vector along the $i$th direction. Therefore, $V(\mathbf{w})$ is a short-range interaction with a characteristic length scale $1/|\mathbf{w}| = a$, where $a=|\mathbf{a}_i|$ and $\mathbf{a}_i$ is the lattice vector. Note $V(\mathbf{w})$ is different from typical long-range interactions in Wannier excitons. The semilocal XC already includes exciton binding between the electron and hole pockets at M and $\Gamma$, respectively, albeit slightly underestimating the screening effect.}

{To quantitatively demonstrate the validity of the semilocal functional in describing the intervalley exciton with momentum $\mathbf{q}=\mathbf{w}$, we compare the linear-response TDDFT results obtained from adiabatic PBE (APBE) and BSE kernels. At $\mathbf{q}=\mathbf{w}$, the APBE kernel yields similar absorption spectra with those from BSE kernel [Supplementary Fig. 3]. Since the BSE kernel is a well-accepted accurate description of excitons~\cite{Sottile2003, Marini2003}, this indicates that the electron-hole exchange effect has been well described in the semilocal XC. Therefore, due to the unique band structures of TiSe$_2$, the semilocal XC yields acceptable excitonic interactions.}

Furthermore, the density functional perturbation theory and molecular dynamics calculations reproduce well experimental phonon spectra~\cite{Weber2011, Singh2017a} and thermal conductivity~\cite{Bianco2015, Hellgren2017}, confirming that DFT can accurately describe the EPC in TiSe$_2$. Thus, the semilocal XC and electron-ion dynamics simulations are suitable for tracking photoexcitation physics of TiSe$_2$.

\begin{figure*}
	\centering
	\includegraphics[width=1.0\linewidth]{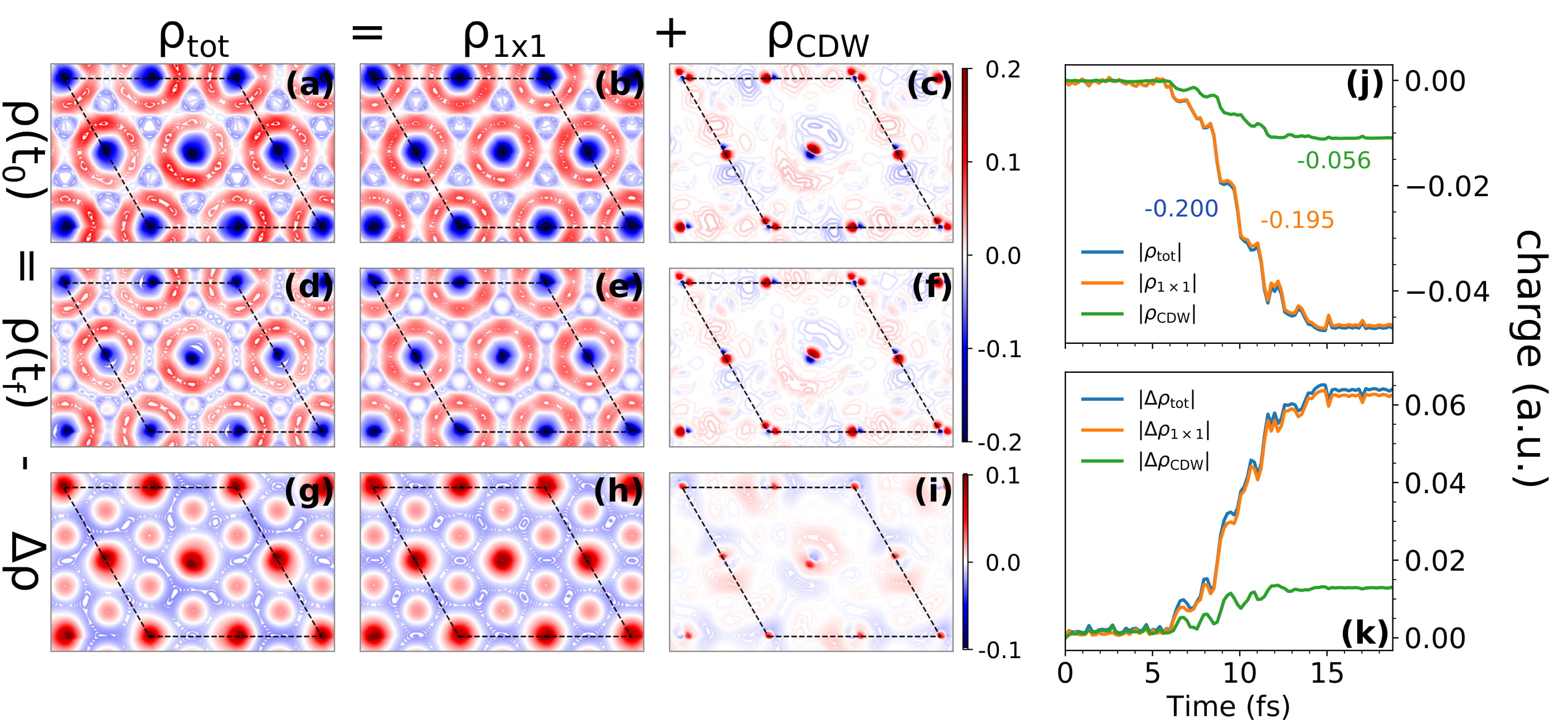}
	\caption{Two-dimensional contour plots for (a) $\rho_{\mathrm{tot}}(x,y,t_0)$, (b) $\rho_{1\times1}(x,y,t_0)$, (c) $\rho_{\mathrm{CDW}}(x,y,t_0)$, (d) $\rho_{\mathrm{tot}}(x,y,t_f)$, (e) $\rho_{1\times1}(x,y,t_f)$, (f) $\rho_{\mathrm{CDW}}(x,y,t_f)$, (g) $\Delta\rho_{\mathrm{tot}}(x,y)$, (h) $\Delta\rho_{1\times1}(x,y)$, (i) $\Delta\rho_{\mathrm{CDW}}(x,y)$, where $t_0=0$ and $t_f=20$ fs. (j) The bonding charge and (k) anti-bonding charge as a function of time. The initial values in (j) are shifted -0.2, -0.195, -0.056 for $\rho_{\mathrm{tot}}$, $\rho_{1\times1}$, $\rho_{\mathrm{CDW}}$, respectively.}
	\label{fig:2Dimage+orderdyn}
\end{figure*}

\noindent\textbf{CDW dynamics.} To identify the dominant mechanism, we try to supply direct proofs by separating the CDW dynamics from the PLD in the time domain. Instead of directly creating CDW at the normal state, we aim at decreasing the CDW order with the fixed PLD. We analyze the differential charge density $\rho_{\mathrm{tot}}(\mathbf{r},t) = \rho_{\mathrm{chg}}(\mathbf{r},t) - \rho_{\mathrm{atom}}(\mathbf{r}, t)$ as a function of time, where $\rho_{\mathrm{chg}}$ is the charge density and $\rho_{\mathrm{atom}}$ is the superposition of the atomic charge densities. Thus, $\rho_{\mathrm{tot}}(\mathbf{r},t)$ features the spacial distribution of the bonding (+) and antibonding (--) densities. We divide  $\rho_{\mathrm{tot}}(\mathbf{r},t)$ into two parts $\rho_{1\times1}(\mathbf{r})$ and $\rho_{\mathrm{CDW}}(\mathbf{r})$, where $\rho_{1\times1}(\mathbf{r}) = [\rho_{\mathrm{tot}}(\mathbf{r} + \mathbf{a}_i) + \rho_{\mathrm{tot}}(\mathbf{r})]/2$, and $\rho_{\mathrm{CDW}}(\mathbf{r}) = \rho_{\mathrm{tot}}(\mathbf{r}) - \rho_{1\times1}(\mathbf{r})$. Obviously, $\rho_{1\times1}(\mathbf{r} + \mathbf{a}_i) = \rho_{1\times1}(\mathbf{r})$ characterizes the original $1\times1$ order, while $\rho_{\mathrm{CDW}}(\mathbf{r})$ characterizes the strength of the CDW order. A laser induced charge difference is characterized by $\Delta\rho_{i}(\mathbf{r}) = \rho_i(\mathbf{r},t_f) - \rho_i(\mathbf{r},t_0)$ ($i=\mathrm{tot}$, $1\times1$, $\mathrm{CDW}$).

Figure~\ref{fig:2Dimage+orderdyn} shows the two-dimensional contour of the charge density $\rho_i(x,y,t)$ ($i=\mathrm{tot}$, $1\times1$, $\mathrm{CDW}$), which is $\rho_i(\mathbf{r},t)$ averaged over $z$ direction. Comparing Fig.~\ref{fig:2Dimage+orderdyn}(a)-(c), we find that $\rho_{1\times1}(x,y,t_0)$ is the main ingredient of $\rho_{\mathrm{tot}}(x,y,t_0)$, even with the presence of PLD, while $\rho_{\mathrm{CDW}}(x,y,t_0)$ is localized around the Ti positions. Starting from the CDW ground state with the fixed PLD, the CDW order $\rho_{\mathrm{CDW}}(x,y,t)$ decreases after the laser illumination: (i) The laser induces electron transfer from the bonding area to the antibonding area. The induced charge $\Delta\rho_\mathrm{tot}(x,y,t)$ [Fig.~\ref{fig:2Dimage+orderdyn}(g)] is opposite to ground state charge $\rho_\mathrm{tot}(x, y, t_0)$ [Fig.~\ref{fig:2Dimage+orderdyn}(a)]. (ii) The majority of the induced charge has the $1\times1$ periodicity, i.e. $\Delta\rho_\mathrm{1\times1}(x,y,t)$ [Fig.~\ref{fig:2Dimage+orderdyn}(h)] dominates $\Delta\rho_\mathrm{tot}(x,y,t)$ [Fig.~\ref{fig:2Dimage+orderdyn}(g)]. (iii) The most important feature is that the induced CDW charge density $\Delta\rho_{\mathrm{CDW}}(x,y)$ [Fig.~\ref{fig:2Dimage+orderdyn}(i)] is opposite to the original $\rho_{\mathrm{CDW}}(x,y,t_0)$ [Fig.~\ref{fig:2Dimage+orderdyn}(c)], indicating a $20\%$ decrease in the excitonic order.

We analyze the integrated charges $Q_{i}(t) = \int |\rho_i(\mathbf{r},t)| d\mathbf{r}$ and $C_{i}(t) = \int |\rho_i(\mathbf{r}, t) - \rho_i(\mathbf{r}, t_0)| d\mathbf{r}$ ($i=\mathrm{tot}$, $1\times1$, $\mathrm{CDW}$) for quantitative comparisons. 
The former characterizes the strength of bonding, while the latter denotes the weakening of bonding states, i.e. the strength of anti-bonding.  
As shown in Fig.~\ref{fig:2Dimage+orderdyn}(j), the percentage of the decrease $[Q_{i}(t_f)-Q_{i}(t_0)]/Q_{i}(t_0)$ is 23.6\%, 23.8\% and 19.7\% for $i=\mathrm{tot}$, $1\times1$, $\mathrm{CDW}$, respectively. Meanwhile, $C_{i}(t)$ increases, with a ratio $C_{\mathrm{tot}}(t_f):C_{\mathrm{1\times1}}(t_f):C_{\mathrm{CDW}}(t_f) = 1:0.98:0.20$, slightly different from the ratio of the initial bonding charges $Q_{\mathrm{tot}}(t_0):Q_{\mathrm{1\times1}}(t_0):Q_{\mathrm{CDW}}(t_0) = 1:0.98:0.27$. Thus, laser-induced bonding-antibonding transfer is nearly homogeneous, lowering both the $1\times1$ and CDW order proportionally. Since the decrease in $|Q_{\mathrm{1\times1},\mathrm{bond}}|$ affects all chemical bonds homogeneously, as an overall effect, the decrease in $|Q_{\mathrm{tot},\mathrm{bond}}|$ lowers the stability of CDW. 

\begin{figure}
	\centering
	\includegraphics[width=1\linewidth]{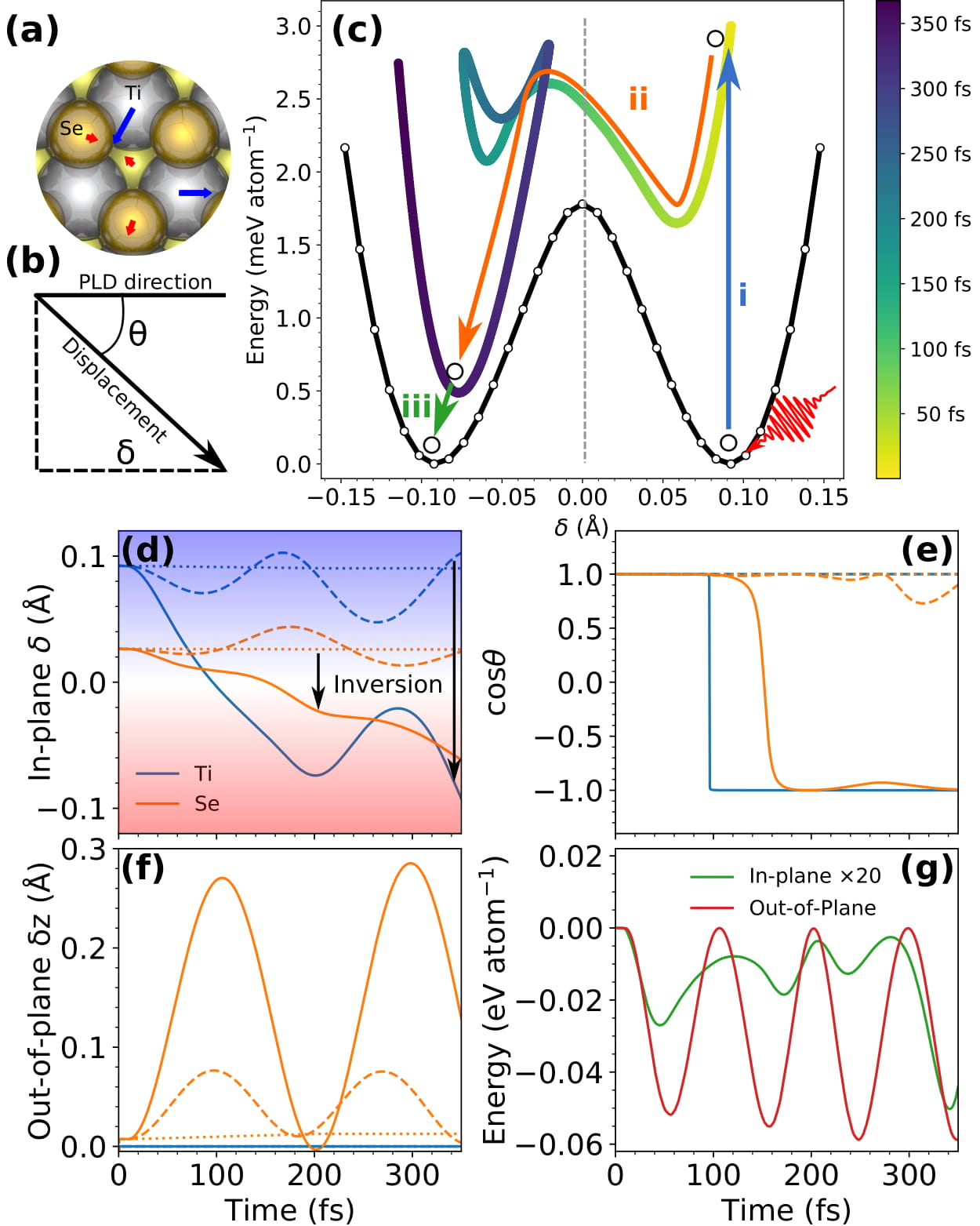}
	\caption{(a) All the symmetrically inequivalent atoms with nontrivial PLD. (b) The illustration of $\delta$ and $\theta$. 
	{(c) The potential energy surfaces (PES). The black and colored lines denote the ground state PES and the non-equilibrium TD-PES $E[\delta(t)]$, respectively.}
	The (d) $\delta(t)$, (e) $\cos\theta(t)$, and (f) $\delta z(t)$ as a function of time. (g) The minus of the kinetic energies of out-of-plane movements $-E_{\mathrm{kin}}(t)$ and in-plane movements $-E_{\mathrm{kout}}(t)$ as a function of time. The solid, dash and dotted lines denote the $I = I_0$, $I = I_0/4$, and the quenched case ($I = I_0$ but with 1.4\% energy dissipation), respectively.}
	\label{fig:Figure4}
\end{figure}

\begin{figure*}
	\centering
	\includegraphics[width=1.0\linewidth]{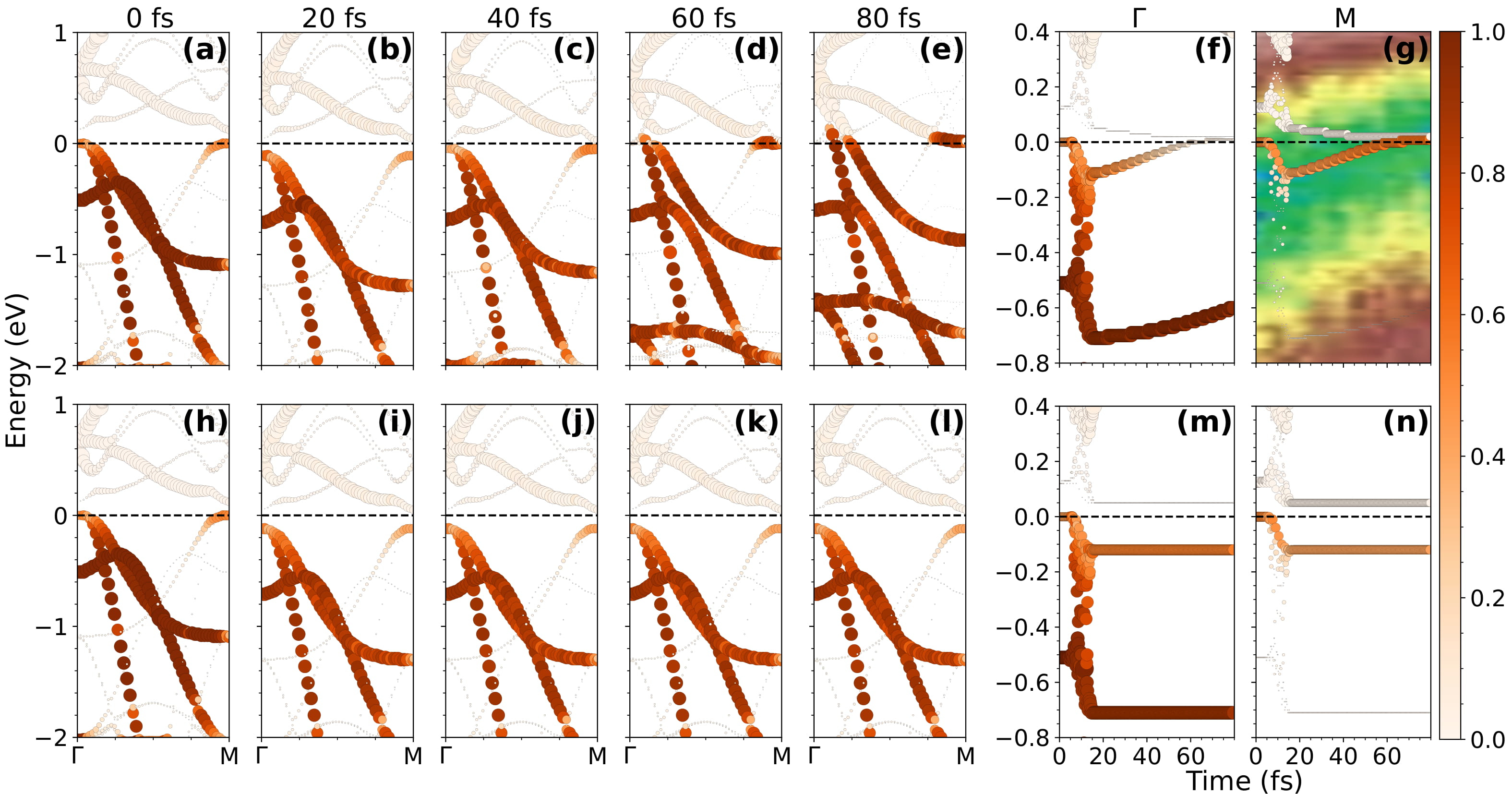}
	\caption{Snapshots of TD-EBS for (a-e) the PLD  dynamics, and (h-l) the quenched case. EBS at $\Gamma$ and M point as a function of time for (f-g) PLD dynamics and (m-n) the quenched case. The color bar denotes the carrier population. The squares in (g) mark the experimental tr-ARPES data, reproduced from Ref.~\cite{Hellmann2012}.
	}
	\label{fig:tdebs}
\end{figure*}

\noindent\textbf{PLD dynamics.} The laser-induced CDW instability would further trigger dynamic changes in PLD, which are simulated from first principles. Among the atoms in the layer of $2\times2$ cell, only one Ti atom and one Se atom are nontrivial symmetry-inequivalent atoms, as shown in Fig.~\ref{fig:Figure4}(a). We present the time-dependent displacement $\mathbf{d}_i(t)$($i = \mathrm{Ti}, \mathrm{Se}$) by the projection of $\mathbf{d}_i(t)$ on the PLD direction $\delta_i(t) = \mathbf{d}_i(t)\cdot\mathbf{d}_i(t_0)/|\mathbf{d}_i(t_0)|$ and the angle $\theta(t)$ between $\mathbf{d}_i(t)$ and $\mathbf{d}_i(t_0)$ [Fig.~\ref{fig:Figure4}(b)]. We find that laser triggers a set of movements $\Delta\mathbf{d}_i(t) = \mathbf{d}_i(t) - \mathbf{d}_i(t_0)$ which are exactly antiparallel to $\mathbf{d}_i(t_0)$, noticing that $\delta_i$ decreases with all $\cos\theta \approx \pm 1$ as shown in Fig.~\ref{fig:Figure4}(d-e). Besides the in-plane movements, the overall bonding-antibonding charge transition introduces an out-of-plane breathing mode between the Se layer and the Ti layer $z_{\mathrm{Se}}(t)$ [Fig.~\ref{fig:Figure4}(f)]. 
{This is the established out-of-plane $A_{1g}$ mode with the periodicity of 175~fs (5.7~THz) when $I=I_0/4$, which is slightly smaller than the experimental value 5.9~THz~\cite{Snow2003,Wang2018g}. Higher laser fluence $I=I_0$ further decreases the frequency to 5.1~THz, due to the weakened Ti-Se bonds by photocarriers.}

We analyze the dynamical potential energy surface (PES) related to the PLD dynamics. To focus on the in-plane PLD movements other than the out-of-plane $A_{1g}$ mode, we plot the minus of the kinetic energies of these two movements $-E_{\mathrm{kin}}$ and $-E_{\mathrm{kout}}$, respectively. Since the total energy $E_{\mathrm{tot}}(t) = E_{\mathrm{p}}(t) + E_{\mathrm{kin}}(t) + E_{\mathrm{kout}}(t)$ is conserved, where $E_{\mathrm{p}}(t)$ is the potential energy, we use $-E_{\mathrm{kin}}(t)$ to characterize the dynamical PES of the PLD. As shown in Fig.~\ref{fig:Figure4}(g), the dynamical potential energy decreases due to the in-plane PLD movements, together with the energy oscillations attributed to the out-of-plane $A_{1g}$ mode. This indicates that the laser induced PLD movements stabilize the system. Within the simulation time, $\mathbf{d}_i(t)$ has been completely inverted $\mathbf{d}_i(t) = -\mathbf{d}_i(t_0)$ at around $300$~fs, lowering the dynamical PES and creating a new quasi-equilibrium state. We note that the inversion of PLD produces a degenerate CDW state which is symmetry-equivalent to the original state. {As shown in Fig.~\ref{fig:Figure4}(c), we generate the dynamical PES by plotting $-E_{\mathrm{kin}}(t)$ as a function of $\delta(t)$.
This implies a scenario of ultrafast PLD switch, which includes three consecutive steps: \textbf{i}) In the first 20~fs, the laser pulses pump the electrons from the bonding state to the anti-bonding state, which leads to the change of the potential energy surface.}

{The calculated double-well PES yields a vibration frequency of 4.3 THz, which is consistent with the oscillation period of the amplitude mode of about 220-240 fs observed in dynamic simulations (Fig. 4d). This frequency is comparable to the experimental value of 3.4 THz observed for the $A_{1g}$ CDW amplitude mode~\cite{Snow2003,Wang2018g}. The difference between theory and experiment is attributed to possible mixing of other phonon modes and inadequate accuracy of semilocal functionals to treat low-energy phonons. The softening of the double-well potential is evidenced by the fact that the CDW mode oscillations disappear when the laser fluence increases from $I_0/4$ to $I_0$.}

\textbf{ii}) The lattice transforms into a structure with the opposite PLD along the non-equilibrium TD-PES during 20--300~fs.
\textbf{\textbf{iii})} The photoexcited system might relax into the ground state after the recombination of photocarriers. This process is beyond our simulations and requires multichannel and large-scale modeling to properly account for the decoherence~\cite{Floss2017, Yamada2018, Floss2019} and dissipation effects~\cite{Tully1990, Gossel2018}. Here, we complete the story by briefly discussing the long-time behaviors: in the experiments, the ionic movements in different unit cells have different phases due to the finite thickness of the material, the inhomogeneous spatial distribution of the laser spot, as well as thermal fluctuations. Via phonon-phonon scattering, the equilibrium temperature gradually forms at the timescale of picoseconds. The thermally driven phase transitions may occur through an increasing population of phonons and changes in the atomic potential via anharmonic phonon interactions. Thus, the nonthermal CDW dynamics observed here is separated from thermal transitions at different time scales.

\noindent\textbf{Dynamic interplay.} The highly directional PLD movements are clearly not the reason but the consequence of the reduced excitonic order. In contrast to the complete PLD inversion for the $I=I_0$ case,  $\mathbf{d}_i(t)$ only oscillates around the original value when laser intensity decreases to $I_0/4$. 
The formation/decay of the CDW/PLD is electronically initialized, which supports the excitonic mechanism.
{This intensity dependence is also experimentally observed~\cite{Mohr-Vorobeva2011, Porer2014}: the $A_{1g}$ CDW mode is softened as the laser fluence increases, which serves as a precursor of the PLD melting.}
Meanwhile, there are extra factors in the following dynamics. Assuming that the excitonic pairing is the only driving force, an inversion of electronic order $\rho_{\mathrm{CDW}}(\mathbf{r}, t_f) = -\rho_{\mathrm{CDW}}(\mathbf{r}, t_0)$ is required to reverse the PLD $\mathbf{d}_i(t) = -\mathbf{d}_i(t_0)$. Considering $\rho_{\mathrm{CDW}}$ only decreases by about 20\%, the PLD should be only perturbed.

We design another numerical experiment by introducing a thermostat and removing a small fraction of kinetic energy at the rate of 0.01~eV atom$^{-1}$ ps$^{-1}$. We note that the kinetic energy $E_{\mathrm{k}}(t)$ increases by 0.66~eV atom$^{-1}$ ps$^{-1}$ within 60~fs, as shown in Fig.~\ref{fig:Figure4}(g). Thus the thermostat, affecting only 1.4\% of the kinetic energy change, is expected to slightly delay the dynamics of PLD $\delta(t)$. However, despite that the dissipation in the kinetic energy is small ($1.4\%$), the PLD dynamics is completely quenched (Fig.~\ref{fig:Figure4}, dotted lines). It strongly implies that the increase in $E_{\mathrm{k}}(t)$ is nonlinear, with a lower rate at the beginning. The fragile initial movement of $\delta(t)$ will be completely quenched with a small dissipation rate. Since the PLD dynamics is sensitively dependent on its own trajectory, we infer that there exists a self-amplified electron-phonon mechanism in the PLD dynamics initiated by exciton binding.


Accordingly, we demonstrate the impact of ionic movements on electronic structure through analyzing TD-EBS along $\Gamma$-M in Fig.~\ref{fig:tdebs}(a-f). The laser pulse first shifts the EBS downwards and enlarges the bandgap; then VBs are raised by $\sim$0.18~eV while CBs barely change during 20-80~fs, in excellent agreement with tr-ARPES data. It leads to the vanishing band gap and forms hole and electron pockets, boosting the relaxation of hole carriers towards the maximum of VBs (VBM). The carrier relaxation lowers the total energy of the system. In comparison, in the quenched case where ions barely move [Fig.~\ref{fig:tdebs}(h-l)], the dynamic electronic evolution -- the gap closing and carrier relaxation -- are completely absent, implying that the measured changes in EBS are highly correlated with ionic movements. We thus identify a self-amplified process from these observations: the decrease in the PLD $\rightarrow$ the decrease in the bandgap $\rightarrow$ the decrease in the energy of carriers $\rightarrow$ the further decrease in PLD. We expect such a dynamic interplay between electronic order and PLD, ignored in previous studies, would sustain in a variety of CDW materials such as NbSe$_2$, TaS$_2$, LaTe$_3$, etc~\cite{Hellgren2017,Zhang2018CDWTaS2}. The results highlight the entangled dynamics between different degree of freedoms in quantum materials.  

\vspace{1cm}\noindent\textbf{Discussion}\\
To complete our story, we note the following facts towards building a unified picture for experimental observations.
\textbf{i}) The effective lattice temperature is relatively low during the whole laser-induced dynamics. The maximum kinetic energy of ions is 0.06 eV atom$^{-1}$ in a temporary period (corresponding to a transient effective temperature of $450$~K). Although the low kinetic energy of ions may not be directly connected to nonthermal processes, we believe the processes discussed above are primarily nonthermal since they are short-lived ($<$300 fs) and the excited-state carriers are clearly present (Fig. 5). In this case the structural soft-mode potential is relaxed or even driven to a single well-potential by excitation of the electronic degree of freedom. This is also consistent with the experimental observations~\cite{Mohr-Vorobeva2011} and previous studies~\cite{Lian2019Silicon, Zhang2018CDWTaS2, Lian2016NTM}.
\textbf{ii}) The CDW order is disturbed instantly by the laser pulse within the timescale of 10~fs, since it is originated from homogeneous electronic transitions. No buildup time in electron-hole screening is needed. It shows that, the $\sim$100~fs shift in tr-ARPES signals~\cite{Hellmann2012} is not caused by the buildup of electron-hole screening, but by the EPC-assisted PLD dynamics $\mathbf{d}_i(t)$. Apparently, the phonon response within 100~fs is contradictory to the well accepted timescale of phonon dynamics $\tau_{\mathrm{ph}} >1$~ps. However, we note that the photo-induced PLD dynamics are coherently driven not by the phonon-phonon scattering, but directly by the electronic transitions. Thus, the response shares the same timescale as the electron-phonon scattering of $\sim$100~fs.
\textbf{iii}) With the laser intensity used in our simulations, the laser pulses induce the PLD dynamics but barely change the intrinsic phonon modes (e.g. out-of-plane $A_{1g}$). This explains the sustainable characteristic peaks of PLD with the melted exciton order as previously observed in Ref.~\cite{Porer2014}.
\textbf{iv}) {There are open discussions in literature on whether excitonic interaction or the Jahn-Teller distortion is dominant for the formation of CDW in TiSe$_2$. Our simulations seem to indicate the two effects could coexist in a short period perturbed by photoexcitation. From Fig.~\ref{fig:2Dimage+orderdyn} and~\ref{fig:Figure4} it is shown that while there is no significant lattice distortion (thus no changes in Jahn-Teller interaction), the electronic order is suppressed by 20\% in the first 20 fs, beaconing excitonic interactions. On the other hand, the following self-amplifying process suggests dynamic Jahn-Teller lattice distortion is crucial for the subsequent decay/formation of CDW state.}

We demonstrate an entangled exciton-phonon mechanism for charge ordering in 1T-TiSe$_2$ by \textit{ab initio} ultrafast rt-TDDFT simulations of photoexcitation dynamics. We show that laser pulses knock down the CDW order via inducing a homogeneous bonding-antibonding electronic transition. The weakened CDW order then triggers ionic movements exactly anti-parallel to the original PLD, but cannot solely drive the observed PLD inversion. Instead, assisted by the EPC, a self-amplification mechanism drives the PLD dynamics initiated by laser excitation, in excellent agreement with the tr-ARPES measurements. We propose that both the excitonic pairing and the electron-phonon coupling contribute to the CDW/PLD formation, but in a different timescale: the excitonic pairing initializes the formation of CDW, while the electron-phonon coupling promotes the following dynamics through a self-amplification process.

\vspace{1 cm}\noindent\textbf{Method}\\ 
Following our previous efforts in developing time dependent \textit{ab initio} package (TDAP)~\cite{Meng2008a,Lian2018MultiK,Lian2018AdvTheo}, we implemented TDDFT algorithm in the plane-wave code \textsf{Quantum Espresso}~\cite{Giannozzi2009}. The details are described in Supplementary Note 1. We used the projector augmented-waves method (PAW)~\cite{Blochl1994} and the Perdew-Burke-Ernzerhof (PBE) exchange-correlation (XC) functional~\cite{Perdew1996} in both DFT and TDDFT calculations. Pseudopotentials were generated using \textsf{pslibrary}~\cite{DalCorso2014}. The plane-wave energy cutoff was set to 55~Ry. Brillouin zone (BZ) was sampled using Monkhorst-Pack scheme~\cite{Monkhorst1976} with a $6\times6\times3$ $k$-point mesh. Band unfolding techniques were utilized to generate the effective band structure (EBS)~\cite{Ku2010, Popescu2012} with a modified version of \textsf{BandUP} code~\cite{Medeiros2014,Medeiros2015,Lian2017SiUnfold}.
{Onsite Coulomb repulsion $U=3.5$~eV was added to the Ti atom to reproduce the experimental band structure, while we used $U=0$ in the dynamic TDDFT calculations and structural optimization.}
The electron timestep $\delta t$ is  $4\times10^{-4}$~a.u.=$0.1935$~attosecond and the ion timestep $\Delta t$ is $0.194$~fs.
Laser pulses with a wavelength $\lambda = 800$~nm, width $\sigma=12$ fs and  fluence $I_0 = 2.1$~mJ cm$^{-2}$ centered at $t=10$ fs is utilized. 
{The linear-response TDDFT calculations were carried out with the YAMBO package~\cite{Marini2009}.}
This setup, which is similar to that used in experiment, reproduces the measured number of excited carriers in TiSe$_2$(Supplementary Fig. 2).

\vspace{1 cm}\noindent\textbf{Acknowledgment}\\
We acknowledge partial financial supports from MOST (grants 2016YFA0300902 and 2015CB921001), NSFC (grants 91850120, 11774396, 11934003), and CAS (XDB07030100).

\vspace{1 cm}\noindent\textbf{Author Contributions}\\
S.M. conceived and directed the research. C.L. performed the calculations and analyzed the data. C.L., S.-J.Z., S.-Q.H., M.-X.G. and S.M. participated in the discussion. C.L. and S.M. wrote the paper with contributions from all the other authors.

\vspace{1 cm}\noindent\textbf{Competing interests}: The authors declare no competing interests.


\begin{thebibliography}{100}
\expandafter\ifx\csname url\endcsname\relax
  \def\url#1{\texttt{#1}}\fi
\expandafter\ifx\csname urlprefix\endcsname\relax\def\urlprefix{URL }\fi
\providecommand{\bibinfo}[2]{#2}
\providecommand{\eprint}[2][]{\url{#2}}

\bibitem{Morosan2006}
\bibinfo{author}{Morosan, E.} \emph{et~al.}
\newblock \bibinfo{title}{Superconductivity in {Cu$_x$TiSe$_2$}}.
\newblock \emph{\bibinfo{journal}{Nat. Phys.}} \textbf{\bibinfo{volume}{2}},
  \bibinfo{pages}{544--550} (\bibinfo{year}{2006}).

\bibitem{Li2007}
\bibinfo{author}{Li, G.} \emph{et~al.}
\newblock \bibinfo{title}{{S}emimetal-to-{S}emimetal {C}harge {D}ensity {W}ave
  {T}ransition in 1{T}-{{T}i{S}e}$_2$}.
\newblock \emph{\bibinfo{journal}{Phys. Rev. Lett.}}
  \textbf{\bibinfo{volume}{99}}, \bibinfo{pages}{027404}
  (\bibinfo{year}{2007}).

\bibitem{Kusmartseva2009}
\bibinfo{author}{Kusmartseva, A.~F.}, \bibinfo{author}{Sipos, B.},
  \bibinfo{author}{Berger, H.}, \bibinfo{author}{Forr{\'{o}}, L.} \&
  \bibinfo{author}{Tuti{\v{s}}, E.}
\newblock \bibinfo{title}{{P}ressure {I}nduced {S}uperconductivity in
  {P}ristine 1{T}-{{T}i{S}e}$_2$}.
\newblock \emph{\bibinfo{journal}{Phys. Rev. Lett.}}
  \textbf{\bibinfo{volume}{103}}, \bibinfo{pages}{236401}
  (\bibinfo{year}{2009}).

\bibitem{Yao2018}
\bibinfo{author}{Yao, Q.} \emph{et~al.}
\newblock \bibinfo{title}{Charge transfer effects in naturally occurring van
  der waals heterostructures ({PbSe})$_{1.16}$({TiSe}$_2$)$_m$(m=1,2)}.
\newblock \emph{\bibinfo{journal}{Phys. Rev. Lett.}}
  \textbf{\bibinfo{volume}{120}}, \bibinfo{pages}{106401}
  (\bibinfo{year}{2018}).

\bibitem{Wei2017b}
\bibinfo{author}{Wei, M.~J.} \emph{et~al.}
\newblock \bibinfo{title}{Manipulating charge density wave order in monolayer
  1{T}-{TiSe}$_2$ by strain and charge doping: {A} first-principles
  investigation}.
\newblock \emph{\bibinfo{journal}{Phys. Rev. B}} \textbf{\bibinfo{volume}{96}},
  \bibinfo{pages}{165404} (\bibinfo{year}{2017}).

\bibitem{Kogar2017a}
\bibinfo{author}{Kogar, A.} \emph{et~al.}
\newblock \bibinfo{title}{Observation of a charge density wave incommensuration
  near the superconducting dome in {{C}u$_x${T}i{S}e}$_2$}.
\newblock \emph{\bibinfo{journal}{Phys. Rev. Lett.}}
  \textbf{\bibinfo{volume}{118}}, \bibinfo{pages}{027002}
  (\bibinfo{year}{2017}).

\bibitem{Yan2017}
\bibinfo{author}{Yan, S.} \emph{et~al.}
\newblock \bibinfo{title}{Influence of domain walls in the incommensurate
  charge density wave state of {C}u intercalated 1{T}-{{T}i{S}e}$_2$}.
\newblock \emph{\bibinfo{journal}{Phys. Rev. Lett.}}
  \textbf{\bibinfo{volume}{118}}, \bibinfo{pages}{106405}
  (\bibinfo{year}{2017}).

\bibitem{Medvecka2016}
\bibinfo{author}{Medveck{\'{a}}, Z.} \emph{et~al.}
\newblock \bibinfo{title}{{O}bservation of a transverse {M}eissner effect {in
  {C}u$_x${T}i{S}e}$_2$ single crystals}.
\newblock \emph{\bibinfo{journal}{Phys. Rev. B}} \textbf{\bibinfo{volume}{93}},
  \bibinfo{pages}{100501} (\bibinfo{year}{2016}).

\bibitem{Das2015}
\bibinfo{author}{Das, T.} \& \bibinfo{author}{Dolui, K.}
\newblock \bibinfo{title}{{S}uperconducting dome {in {M}o{S}}$_2$ and
  {T}i{S}e$_2$ generated by quasiparticle-phonon coupling}.
\newblock \emph{\bibinfo{journal}{Phys. Rev. B}} \textbf{\bibinfo{volume}{91}},
  \bibinfo{pages}{094510} (\bibinfo{year}{2015}).

\bibitem{Luna2015}
\bibinfo{author}{Luna, K.}, \bibinfo{author}{Wu, P.~M.}, \bibinfo{author}{Chen,
  J.~S.}, \bibinfo{author}{Morosan, E.} \& \bibinfo{author}{Beasley, M.~R.}
\newblock \bibinfo{title}{{P}oint-contact tunneling spectroscopy measurement
  {of {C}u$_x${T}i{S}e$_2$}: {D}isorder-enhanced {C}oulomb effects}.
\newblock \emph{\bibinfo{journal}{Phys. Rev. B}} \textbf{\bibinfo{volume}{91}},
  \bibinfo{pages}{094509} (\bibinfo{year}{2015}).

\bibitem{Ganesh2014}
\bibinfo{author}{Ganesh, R.}, \bibinfo{author}{Baskaran, G.},
  \bibinfo{author}{van~den Brink, J.} \& \bibinfo{author}{Efremov, D.~V.}
\newblock \bibinfo{title}{{T}heoretical {P}rediction of a {T}ime-{R}eversal
  {B}roken {C}hiral {S}uperconducting {P}hase {D}riven by {E}lectronic
  {C}orrelations in a {{S}ingle {T}i{S}e}$_2$ {L}ayer}.
\newblock \emph{\bibinfo{journal}{Phys. Rev. Lett.}}
  \textbf{\bibinfo{volume}{113}}, \bibinfo{pages}{177001}
  (\bibinfo{year}{2014}).

\bibitem{Joe2014}
\bibinfo{author}{Joe, Y.~I.} \emph{et~al.}
\newblock \bibinfo{title}{Emergence of charge density wave domain walls above
  the superconducting dome in 1{T}-{TiSe}$_2$}.
\newblock \emph{\bibinfo{journal}{Nat. Phys.}} \textbf{\bibinfo{volume}{10}},
  \bibinfo{pages}{421} (\bibinfo{year}{2014}).

\bibitem{Kacmarcik2013}
\bibinfo{author}{Ka{\v{c}}mar{\v{c}}{\'{\i}}k, J.} \emph{et~al.}
\newblock \bibinfo{title}{{H}eat capacity of single-crystal
  {{C}u$_x${T}i{S}e$_2$} superconductors}.
\newblock \emph{\bibinfo{journal}{Phys. Rev. B}} \textbf{\bibinfo{volume}{88}},
  \bibinfo{pages}{020507} (\bibinfo{year}{2013}).

\bibitem{Husanikova2013}
\bibinfo{author}{Husan{\'{\i}}kov{\'{a}}, P.} \emph{et~al.}
\newblock \bibinfo{title}{{M}agnetization properties and vortex phase diagram
  of {{C}u$_x${T}i{S}e}$_2$ single crystals}.
\newblock \emph{\bibinfo{journal}{Phys. Rev. B}} \textbf{\bibinfo{volume}{88}},
  \bibinfo{pages}{174501} (\bibinfo{year}{2013}).

\bibitem{Iavarone2012}
\bibinfo{author}{Iavarone, M.} \emph{et~al.}
\newblock \bibinfo{title}{{E}volution of the charge density wave state in
  {C}u$_x${T}i{S}e$_2$}.
\newblock \emph{\bibinfo{journal}{Phys. Rev. B}} \textbf{\bibinfo{volume}{85}},
  \bibinfo{pages}{155103} (\bibinfo{year}{2012}).

\bibitem{Jeong2010}
\bibinfo{author}{Jeong, J.}, \bibinfo{author}{Jeong, J.}, \bibinfo{author}{Noh,
  H.-J.}, \bibinfo{author}{Kim, S.~B.} \& \bibinfo{author}{Kim, H.-D.}
\newblock \bibinfo{title}{{E}lectronic structure study of {C}u-doped
  1{T}-{{T}i{S}e}$_2$ by angle-resolved photoemission spectroscopy}.
\newblock \emph{\bibinfo{journal}{Physica C}} \textbf{\bibinfo{volume}{470}},
  \bibinfo{pages}{S648} (\bibinfo{year}{2010}).

\bibitem{Zaberchik2010}
\bibinfo{author}{Zaberchik, M.} \emph{et~al.}
\newblock \bibinfo{title}{{P}ossible evidence of a two-gap structure for {the
  {C}u$_x${T}i{S}e}$_2$ superconductor}.
\newblock \emph{\bibinfo{journal}{Phys. Rev. B}} \textbf{\bibinfo{volume}{81}},
  \bibinfo{pages}{220505} (\bibinfo{year}{2010}).

\bibitem{Giang2010}
\bibinfo{author}{Giang, N.} \emph{et~al.}
\newblock \bibinfo{title}{{S}uperconductivity at 2.3 {K} in the misfit
  compound({{P}b{S}e})$_{1.16}$({{T}i{S}e}$_2$)$_2$}.
\newblock \emph{\bibinfo{journal}{Phys. Rev. B}} \textbf{\bibinfo{volume}{82}},
  \bibinfo{pages}{024503} (\bibinfo{year}{2010}).

\bibitem{Morosan2010}
\bibinfo{author}{Morosan, E.} \emph{et~al.}
\newblock \bibinfo{title}{{M}ultiple electronic transitions and
  superconductivity {in {P}d$_x${T}i{S}e}$_2$}.
\newblock \emph{\bibinfo{journal}{Phys. Rev. B}} \textbf{\bibinfo{volume}{81}},
  \bibinfo{pages}{094524} (\bibinfo{year}{2010}).

\bibitem{Hillier2010}
\bibinfo{author}{Hillier, A.~D.} \emph{et~al.}
\newblock \bibinfo{title}{{P}robing the superconducting ground state near the
  charge density wave phase transition {in {C}u}$_{0.06}${T}i{S}e$_2$}.
\newblock \emph{\bibinfo{journal}{Phys. Rev. B}} \textbf{\bibinfo{volume}{81}},
  \bibinfo{pages}{092507} (\bibinfo{year}{2010}).

\bibitem{Barath2008}
\bibinfo{author}{Barath, H.} \emph{et~al.}
\newblock \bibinfo{title}{{Q}uantum and {C}lassical {M}ode {S}oftening {N}ear
  the {C}harge-{D}ensity-{W}ave{\textendash}{S}uperconductor {T}ransition {of
  {C}u$_x${T}i{S}e$_2$}}.
\newblock \emph{\bibinfo{journal}{Phys. Rev. Lett.}}
  \textbf{\bibinfo{volume}{100}}, \bibinfo{pages}{106402}
  (\bibinfo{year}{2008}).

\bibitem{Li2007b}
\bibinfo{author}{Li, S.~Y.}, \bibinfo{author}{Wu, G.}, \bibinfo{author}{Chen,
  X.~H.} \& \bibinfo{author}{Taillefer, L.}
\newblock \bibinfo{title}{{S}ingle-{G}aps-{W}ave {S}uperconductivity near the
  {C}harge-{D}ensity-{W}ave {Q}uantum {C}ritical {P}oint {in
  {C}u$_x${T}i{S}e}$_2$}.
\newblock \emph{\bibinfo{journal}{Phys. Rev. Lett.}}
  \textbf{\bibinfo{volume}{99}}, \bibinfo{pages}{107001}
  (\bibinfo{year}{2007}).

\bibitem{Cui2006}
\bibinfo{author}{Cui, X.~Y.} \emph{et~al.}
\newblock \bibinfo{title}{{D}irect evidence of band modification and
  suppression of superstructure {in {T}i{S}e}$_2$ upon {F}e intercalation: {A}n
  angle-resolved photoemission study}.
\newblock \emph{\bibinfo{journal}{Phys. Rev. B}} \textbf{\bibinfo{volume}{73}},
  \bibinfo{pages}{085111} (\bibinfo{year}{2006}).

\bibitem{Kohn1967}
\bibinfo{author}{Kohn, W.}
\newblock \bibinfo{title}{{E}xcitonic {P}hases}.
\newblock \emph{\bibinfo{journal}{Phys. Rev. Lett.}}
  \textbf{\bibinfo{volume}{19}}, \bibinfo{pages}{439} (\bibinfo{year}{1967}).

\bibitem{Jerome1967}
\bibinfo{author}{J{\'{e}}rome, D.}, \bibinfo{author}{Rice, T.~M.} \&
  \bibinfo{author}{Kohn, W.}
\newblock \bibinfo{title}{{E}xcitonic {I}nsulator}.
\newblock \emph{\bibinfo{journal}{Phys. Rev.}} \textbf{\bibinfo{volume}{158}},
  \bibinfo{pages}{462} (\bibinfo{year}{1967}).

\bibitem{Halperin1968}
\bibinfo{author}{Halperin, B.~I.} \& \bibinfo{author}{Rice, T.~M.}
\newblock \bibinfo{title}{{P}ossible {A}nomalies at a
  {S}emimetal-{S}emiconductor {T}ransistion}.
\newblock \emph{\bibinfo{journal}{Rev. Mod. Phys.}}
  \textbf{\bibinfo{volume}{40}}, \bibinfo{pages}{755} (\bibinfo{year}{1968}).

\bibitem{Rossnagel2002}
\bibinfo{author}{Rossnagel, K.}, \bibinfo{author}{Kipp, L.} \&
  \bibinfo{author}{Skibowski, M.}
\newblock \bibinfo{title}{{C}harge-density-wave phase transition in
  1{T}-{{T}i{S}e}$_2$:{\hspace{0.6em}}{\hspace{0.6em}}{E}xcitonic insulator
  versus band-type {J}ahn-{T}eller mechanism}.
\newblock \emph{\bibinfo{journal}{Phys. Rev. B}} \textbf{\bibinfo{volume}{65}},
  \bibinfo{pages}{235101} (\bibinfo{year}{2002}).

\bibitem{Cercellier2007}
\bibinfo{author}{Cercellier, H.} \emph{et~al.}
\newblock \bibinfo{title}{{E}vidence for an {E}xcitonic {I}nsulator {P}hase in
  1{T}-{{T}i{S}e}$_2$}.
\newblock \emph{\bibinfo{journal}{Phys. Rev. Lett.}}
  \textbf{\bibinfo{volume}{99}}, \bibinfo{pages}{146403}
  (\bibinfo{year}{2007}).

\bibitem{Hughes1977}
\bibinfo{author}{Hughes, H.~P.}
\newblock \bibinfo{title}{{S}tructural distortion in {{T}i{S}e}$_2$and related
  materials-a possible {J}ahn-{T}eller effect?}
\newblock \emph{\bibinfo{journal}{J. Phys. C Solid State}}
  \textbf{\bibinfo{volume}{10}}, \bibinfo{pages}{L319} (\bibinfo{year}{1977}).

\bibitem{Wakabayashi1978}
\bibinfo{author}{Wakabayashi, N.}, \bibinfo{author}{Smith, H.},
  \bibinfo{author}{Woo, K.} \& \bibinfo{author}{Brown, F.}
\newblock \bibinfo{title}{{P}honons and charge density waves in
  1{T}-{{T}i{S}e}$_2$}.
\newblock \emph{\bibinfo{journal}{Solid State Commun.}}
  \textbf{\bibinfo{volume}{28}}, \bibinfo{pages}{923} (\bibinfo{year}{1978}).

\bibitem{Gaby1981}
\bibinfo{author}{Gaby, J.~H.}, \bibinfo{author}{DeLong, B.},
  \bibinfo{author}{Brown, F.}, \bibinfo{author}{Kirby, R.} \&
  \bibinfo{author}{L{\'{e}}vy, F.}
\newblock \bibinfo{title}{{O}rigin of the structural transition in
  {{T}i{S}e}$_2$}.
\newblock \emph{\bibinfo{journal}{Solid State Commun.}}
  \textbf{\bibinfo{volume}{39}}, \bibinfo{pages}{1167} (\bibinfo{year}{1981}).

\bibitem{Motizuki1981}
\bibinfo{author}{Motizuki, K.}, \bibinfo{author}{Suzuki, N.},
  \bibinfo{author}{Yoshida, Y.} \& \bibinfo{author}{Takaoka, Y.}
\newblock \bibinfo{title}{{R}ole of electron-lattice interaction in lattice
  dynamics and lattice instability of 1{T}-{{T}i{S}e}$_2$}.
\newblock \emph{\bibinfo{journal}{Solid State Commun.}}
  \textbf{\bibinfo{volume}{40}}, \bibinfo{pages}{995} (\bibinfo{year}{1981}).

\bibitem{Lopez-Castillo1987}
\bibinfo{author}{Lopez-Castillo, J.~M.} \emph{et~al.}
\newblock \bibinfo{title}{{P}honon-drag effect {in {T}i{S}e}$_{2-x}${S}$_x$
  mixed compounds}.
\newblock \emph{\bibinfo{journal}{Phys. Rev. B}} \textbf{\bibinfo{volume}{36}},
  \bibinfo{pages}{4249} (\bibinfo{year}{1987}).

\bibitem{Holt2001}
\bibinfo{author}{Holt, M.}, \bibinfo{author}{Zschack, P.},
  \bibinfo{author}{Hong, H.}, \bibinfo{author}{Chou, M.~Y.} \&
  \bibinfo{author}{Chiang, T.-C.}
\newblock \bibinfo{title}{{X}-{R}ay {S}tudies of {P}honon {S}oftening {in
  {T}i{S}e}$_2$}.
\newblock \emph{\bibinfo{journal}{Phys. Rev. Lett.}}
  \textbf{\bibinfo{volume}{86}}, \bibinfo{pages}{3799} (\bibinfo{year}{2001}).

\bibitem{Bussmann-Holder2009}
\bibinfo{author}{Bussmann-Holder, A.} \& \bibinfo{author}{Bishop, A.~R.}
\newblock \bibinfo{title}{{S}uppression of charge-density formation {in
  {T}i{S}e}$_2$ by {C}u doping}.
\newblock \emph{\bibinfo{journal}{Phys. Rev. B}} \textbf{\bibinfo{volume}{79}},
  \bibinfo{pages}{024302} (\bibinfo{year}{2009}).

\bibitem{VanWezel2010a}
\bibinfo{author}{van Wezel, J.}, \bibinfo{author}{Nahai-Williamson, P.} \&
  \bibinfo{author}{Saxena, S.~S.}
\newblock \bibinfo{title}{{E}xciton-phonon-driven charge density wave in
  {{T}i{S}e}$_2$}.
\newblock \emph{\bibinfo{journal}{Phys. Rev. B}} \textbf{\bibinfo{volume}{81}},
  \bibinfo{pages}{165109} (\bibinfo{year}{2010}).

\bibitem{Rossnagel2010}
\bibinfo{author}{Rossnagel, K.}
\newblock \bibinfo{title}{{S}uppression and emergence of charge-density waves
  at the surfaces of layered 1{T}-{{T}i{S}e}$_2$ and 1{T}-{{T}a{S}}$_2$ by in
  {situ{R}b} deposition}.
\newblock \emph{\bibinfo{journal}{New J. Phys.}} \textbf{\bibinfo{volume}{12}},
  \bibinfo{pages}{125018} (\bibinfo{year}{2010}).

\bibitem{Calandra2011}
\bibinfo{author}{Calandra, M.} \& \bibinfo{author}{Mauri, F.}
\newblock \bibinfo{title}{{C}harge-{D}ensity {W}ave and {S}uperconducting
  {D}ome {in {T}i{S}e}$_2$ from {E}lectron-{P}honon {I}nteraction}.
\newblock \emph{\bibinfo{journal}{Phys. Rev. Lett.}}
  \textbf{\bibinfo{volume}{106}}, \bibinfo{pages}{196406}
  (\bibinfo{year}{2011}).

\bibitem{Zhu2012}
\bibinfo{author}{Zhu, Z.}, \bibinfo{author}{Cheng, Y.} \&
  \bibinfo{author}{Schwingenschlögl, U.}
\newblock \bibinfo{title}{{O}rigin of the charge density wave in
  1{T}-{{T}i{S}e}$_2$}.
\newblock \emph{\bibinfo{journal}{Phys. Rev. B}} \textbf{\bibinfo{volume}{85}},
  \bibinfo{pages}{245133} (\bibinfo{year}{2012}).

\bibitem{Zenker2013}
\bibinfo{author}{Zenker, B.}, \bibinfo{author}{Fehske, H.},
  \bibinfo{author}{Beck, H.}, \bibinfo{author}{Monney, C.} \&
  \bibinfo{author}{Bishop, A.~R.}
\newblock \bibinfo{title}{{C}hiral charge order in 1{T}-{{T}i{S}e}$_2$:
  {I}mportance of lattice degrees of freedom}.
\newblock \emph{\bibinfo{journal}{Phys. Rev. B}} \textbf{\bibinfo{volume}{88}},
  \bibinfo{pages}{075138} (\bibinfo{year}{2013}).

\bibitem{Kidd2002}
\bibinfo{author}{Kidd, T.~E.}, \bibinfo{author}{Miller, T.},
  \bibinfo{author}{Chou, M.~Y.} \& \bibinfo{author}{Chiang, T.-C.}
\newblock \bibinfo{title}{{E}lectron-{H}ole {C}oupling and the {C}harge
  {D}ensity {W}ave {T}ransition {in {T}i{S}e}$_2$}.
\newblock \emph{\bibinfo{journal}{Phys. Rev. Lett.}}
  \textbf{\bibinfo{volume}{88}}, \bibinfo{pages}{226402}
  (\bibinfo{year}{2002}).

\bibitem{Monney2012a}
\bibinfo{author}{Monney, C.}, \bibinfo{author}{Monney, G.},
  \bibinfo{author}{Aebi, P.} \& \bibinfo{author}{Beck, H.}
\newblock \bibinfo{title}{{E}lectron{\textendash}hole instability in
  1{T}-{{T}i{S}e}$_2$}.
\newblock \emph{\bibinfo{journal}{New J. Phys.}} \textbf{\bibinfo{volume}{14}},
  \bibinfo{pages}{075026} (\bibinfo{year}{2012}).

\bibitem{VanWezel2010}
\bibinfo{author}{van Wezel, J.}, \bibinfo{author}{Nahai-Williamson, P.} \&
  \bibinfo{author}{Saxena, S.~S.}
\newblock \bibinfo{title}{{A}n alternative interpretation of recent
  {{A}{R}{P}{E}{S}} measurements on {{T}i{S}e}$_2$}.
\newblock \emph{\bibinfo{journal}{EPL-Europhys. Lett.}}
  \textbf{\bibinfo{volume}{89}}, \bibinfo{pages}{47004} (\bibinfo{year}{2010}).

\bibitem{Stoffel1982}
\bibinfo{author}{Stoffel, N.}, \bibinfo{author}{L{\'{e}}vy, F.},
  \bibinfo{author}{Bertoni, C.} \& \bibinfo{author}{Margaritondo, G.}
\newblock \bibinfo{title}{{D}irect evidence for d-band involvement in the
  {{T}i{S}e}$_2$ phase transition}.
\newblock \emph{\bibinfo{journal}{Solid State Commun.}}
  \textbf{\bibinfo{volume}{41}}, \bibinfo{pages}{53} (\bibinfo{year}{1982}).

\bibitem{Monney2015}
\bibinfo{author}{Monney, G.}, \bibinfo{author}{Monney, C.},
  \bibinfo{author}{Hildebrand, B.}, \bibinfo{author}{Aebi, P.} \&
  \bibinfo{author}{Beck, H.}
\newblock \bibinfo{title}{{I}mpact of {E}lectron-{H}ole {C}orrelations on the
  1{T}-{{T}i{S}e}$_2$ {E}lectronic {S}tructure}.
\newblock \emph{\bibinfo{journal}{Phys. Rev. Lett.}}
  \textbf{\bibinfo{volume}{114}}, \bibinfo{pages}{086402}
  (\bibinfo{year}{2015}).

\bibitem{Cazzaniga2012}
\bibinfo{author}{Cazzaniga, M.} \emph{et~al.}
\newblock \bibinfo{title}{{A}b initiomany-body effects in {{T}i{S}e}$_2$: {A}
  possible excitonic insulator scenario from {{G}{W}} band-shape
  renormalization}.
\newblock \emph{\bibinfo{journal}{Phys. Rev. B}} \textbf{\bibinfo{volume}{85}},
  \bibinfo{pages}{195111} (\bibinfo{year}{2012}).

\bibitem{Monney2010}
\bibinfo{author}{Monney, C.} \emph{et~al.}
\newblock \bibinfo{title}{{T}emperature-dependent photoemission on
  1{T}-{{T}i{S}e}$_2$: {I}nterpretation within the exciton condensate phase
  model}.
\newblock \emph{\bibinfo{journal}{Phys. Rev. B}} \textbf{\bibinfo{volume}{81}},
  \bibinfo{pages}{155104} (\bibinfo{year}{2010}).

\bibitem{Anderson1985a}
\bibinfo{author}{Anderson, O.}, \bibinfo{author}{Manzke, R.} \&
  \bibinfo{author}{Skibowski, M.}
\newblock \bibinfo{title}{{T}hree-{D}imensional and {R}elativistic {E}ffects in
  {L}ayered 1{T}-{{T}i{S}e}$_2$}.
\newblock \emph{\bibinfo{journal}{Phys. Rev. Lett.}}
  \textbf{\bibinfo{volume}{55}}, \bibinfo{pages}{2188} (\bibinfo{year}{1985}).

\bibitem{Peng2015b}
\bibinfo{author}{Peng, J.-P.} \emph{et~al.}
\newblock \bibinfo{title}{{M}olecular beam epitaxy growth and scanning
  tunneling microscopy study {of {T}i{S}e}$_2$ ultrathin films}.
\newblock \emph{\bibinfo{journal}{Phys. Rev. B}} \textbf{\bibinfo{volume}{91}},
  \bibinfo{pages}{121113} (\bibinfo{year}{2015}).

\bibitem{Koley2014}
\bibinfo{author}{Koley, S.}, \bibinfo{author}{Laad, M.~S.},
  \bibinfo{author}{Vidhyadhiraja, N.~S.} \& \bibinfo{author}{Taraphder, A.}
\newblock \bibinfo{title}{{P}reformed excitons, orbital selectivity, and charge
  density wave order in 1{T}-{{T}i{S}e}$_2$}.
\newblock \emph{\bibinfo{journal}{Phys. Rev. B}} \textbf{\bibinfo{volume}{90}},
  \bibinfo{pages}{115146} (\bibinfo{year}{2014}).

\bibitem{Novello2017}
\bibinfo{author}{Novello, A.} \emph{et~al.}
\newblock \bibinfo{title}{Stripe and short range order in the charge density
  wave of 1{T}-{{C}u$_x${T}i{S}e}$_2$}.
\newblock \emph{\bibinfo{journal}{Phys. Rev. Lett.}}
  \textbf{\bibinfo{volume}{118}}, \bibinfo{pages}{017002}
  (\bibinfo{year}{2017}).

\bibitem{Monney2012}
\bibinfo{author}{Monney, C.}, \bibinfo{author}{Monney, G.},
  \bibinfo{author}{Aebi, P.} \& \bibinfo{author}{Beck, H.}
\newblock \bibinfo{title}{{E}lectron-hole fluctuation phase in
  1{T}-{{T}i{S}e}$_2$}.
\newblock \emph{\bibinfo{journal}{Phys. Rev. B}} \textbf{\bibinfo{volume}{85}},
  \bibinfo{pages}{235150} (\bibinfo{year}{2012}).

\bibitem{Monney2010a}
\bibinfo{author}{Monney, C.} \emph{et~al.}
\newblock \bibinfo{title}{{P}robing the exciton condensate phase in
  1{T}-{{T}i{S}e}$_2$with photoemission}.
\newblock \emph{\bibinfo{journal}{New J. Phys.}} \textbf{\bibinfo{volume}{12}},
  \bibinfo{pages}{125019} (\bibinfo{year}{2010}).

\bibitem{Monney2011}
\bibinfo{author}{Monney, C.}, \bibinfo{author}{Battaglia, C.},
  \bibinfo{author}{Cercellier, H.}, \bibinfo{author}{Aebi, P.} \&
  \bibinfo{author}{Beck, H.}
\newblock \bibinfo{title}{{E}xciton {C}ondensation {D}riving the {P}eriodic
  {L}attice {D}istortion of 1{T}-{{T}i{S}e}$_2$}.
\newblock \emph{\bibinfo{journal}{Phys. Rev. Lett.}}
  \textbf{\bibinfo{volume}{106}}, \bibinfo{pages}{106404}
  (\bibinfo{year}{2011}).

\bibitem{Monney2009}
\bibinfo{author}{Monney, C.} \emph{et~al.}
\newblock \bibinfo{title}{{S}pontaneous exciton condensation in
  1{T}-{{T}i{S}e}$_2$: {{B}{C}{S}}-like approach}.
\newblock \emph{\bibinfo{journal}{Phys. Rev. B}} \textbf{\bibinfo{volume}{79}},
  \bibinfo{pages}{045116} (\bibinfo{year}{2009}).

\bibitem{Sugawara2016}
\bibinfo{author}{Sugawara, K.} \emph{et~al.}
\newblock \bibinfo{title}{{U}nconventional {C}harge-{D}ensity-{W}ave
  {T}ransition in {M}onolayer 1{T}-{{T}i{S}e}$_2$}.
\newblock \emph{\bibinfo{journal}{{ACS} Nano}} \textbf{\bibinfo{volume}{10}},
  \bibinfo{pages}{1341} (\bibinfo{year}{2015}).

\bibitem{Hildebrand2016a}
\bibinfo{author}{Hildebrand, B.} \emph{et~al.}
\newblock \bibinfo{title}{{S}hort-range phase coherence and origin of the
  1{T}-{{T}i{S}e}$_2$ charge density wave}.
\newblock \emph{\bibinfo{journal}{Phys. Rev. B}} \textbf{\bibinfo{volume}{93}},
  \bibinfo{pages}{125140} (\bibinfo{year}{2016}).

\bibitem{Watanabe2015}
\bibinfo{author}{Watanabe, H.}, \bibinfo{author}{Seki, K.} \&
  \bibinfo{author}{Yunoki, S.}
\newblock \bibinfo{title}{{C}harge-density wave induced by combined
  electron-electron and electron-phonon interactions in 1{T}-{{T}i{S}e}$_2$:
  {A} variational {M}onte {C}arlo study}.
\newblock \emph{\bibinfo{journal}{Phys. Rev. B}} \textbf{\bibinfo{volume}{91}},
  \bibinfo{pages}{205135} (\bibinfo{year}{2015}).

\bibitem{Pillo2000}
\bibinfo{author}{Pillo, T.} \emph{et~al.}
\newblock \bibinfo{title}{{P}hotoemission of bands above the {F}ermi level:
  {T}he excitonic insulator phase transition in 1{T}-{{T}i{S}e}$_2$}.
\newblock \emph{\bibinfo{journal}{Phys. Rev. B}} \textbf{\bibinfo{volume}{61}},
  \bibinfo{pages}{16213} (\bibinfo{year}{2000}).

\bibitem{May2011}
\bibinfo{author}{May, M.~M.}, \bibinfo{author}{Brabetz, C.},
  \bibinfo{author}{Janowitz, C.} \& \bibinfo{author}{Manzke, R.}
\newblock \bibinfo{title}{{C}harge-{D}ensity-{W}ave {P}hase of
  1{T}-{{T}i{S}e}$_2$: {T}he {I}nfluence of {C}onduction {B}and {P}opulation}.
\newblock \emph{\bibinfo{journal}{Phys. Rev. Lett.}}
  \textbf{\bibinfo{volume}{107}}, \bibinfo{pages}{176405}
  (\bibinfo{year}{2011}).

\bibitem{Chen2018c}
\bibinfo{author}{Chen, C.}, \bibinfo{author}{Singh, B.}, \bibinfo{author}{Lin,
  H.} \& \bibinfo{author}{Pereira, V.~M.}
\newblock \bibinfo{title}{Reproduction of the charge density wave phase diagram
  in 1{T}-{TiSe}$_2$ {E}xposes its {E}xcitonic {C}haracter}.
\newblock \emph{\bibinfo{journal}{Phys. Rev. Lett.}}
  \textbf{\bibinfo{volume}{121}}, \bibinfo{pages}{226602}
  (\bibinfo{year}{2018}).

\bibitem{Sundaram2002}
\bibinfo{author}{Sundaram, S.~K.} \& \bibinfo{author}{Mazur, E.}
\newblock \bibinfo{title}{{I}nducing and probing non-thermal transitions in
  semiconductors using femtosecond laser pulses}.
\newblock \emph{\bibinfo{journal}{Nat. Mater.}} \textbf{\bibinfo{volume}{1}},
  \bibinfo{pages}{217} (\bibinfo{year}{2002}).

\bibitem{Rohwer2011a}
\bibinfo{author}{Rohwer, T.} \emph{et~al.}
\newblock \bibinfo{title}{{C}ollapse of long-range charge order tracked by
  time-resolved photoemission at high momenta}.
\newblock \emph{\bibinfo{journal}{Nature}} \textbf{\bibinfo{volume}{471}},
  \bibinfo{pages}{490} (\bibinfo{year}{2011}).

\bibitem{Mathias2016}
\bibinfo{author}{Mathias, S.} \emph{et~al.}
\newblock \bibinfo{title}{{S}elf-amplified photo-induced gap quenching in a
  correlated electron material}.
\newblock \emph{\bibinfo{journal}{Nat. Commun.}} \textbf{\bibinfo{volume}{7}},
  \bibinfo{pages}{12902} (\bibinfo{year}{2016}).

\bibitem{Mohr-Vorobeva2011}
\bibinfo{author}{Möhr-Vorobeva, E.} \emph{et~al.}
\newblock \bibinfo{title}{{N}onthermal {M}elting of a {C}harge {D}ensity {W}ave
  {in {T}i{S}e}$_2$}.
\newblock \emph{\bibinfo{journal}{Phys. Rev. Lett.}}
  \textbf{\bibinfo{volume}{107}}, \bibinfo{pages}{036403}
  (\bibinfo{year}{2011}).

\bibitem{Hellmann2012}
\bibinfo{author}{Hellmann, S.} \emph{et~al.}
\newblock \bibinfo{title}{{T}ime-domain classification of charge-density-wave
  insulators}.
\newblock \emph{\bibinfo{journal}{Nat. Commun.}} \textbf{\bibinfo{volume}{3}},
  \bibinfo{pages}{1069} (\bibinfo{year}{2012}).

\bibitem{Porer2014}
\bibinfo{author}{Porer, M.} \emph{et~al.}
\newblock \bibinfo{title}{{N}on-thermal separation of electronic and structural
  orders in a persisting charge density wave}.
\newblock \emph{\bibinfo{journal}{Nat. Mater.}} \textbf{\bibinfo{volume}{13}},
  \bibinfo{pages}{857} (\bibinfo{year}{2014}).

\bibitem{Runge1984}
\bibinfo{author}{Runge, E.} \& \bibinfo{author}{Gross, E. K.~U.}
\newblock \bibinfo{title}{{D}ensity-{F}unctional {T}heory for
  {T}ime-{D}ependent {S}ystems}.
\newblock \emph{\bibinfo{journal}{Phys. Rev. Lett.}}
  \textbf{\bibinfo{volume}{52}}, \bibinfo{pages}{997} (\bibinfo{year}{1984}).

\bibitem{Bertsch2000}
\bibinfo{author}{Bertsch, G.~F.}, \bibinfo{author}{Iwata, J.-I.},
  \bibinfo{author}{Rubio, A.} \& \bibinfo{author}{Yabana, K.}
\newblock \bibinfo{title}{{R}eal-space, real-time method for the dielectric
  function}.
\newblock \emph{\bibinfo{journal}{Phys. Rev. B}} \textbf{\bibinfo{volume}{62}},
  \bibinfo{pages}{7998} (\bibinfo{year}{2000}).

\bibitem{Yabana2006}
\bibinfo{author}{Yabana, K.}, \bibinfo{author}{Nakatsukasa, T.},
  \bibinfo{author}{Iwata, J.-I.} \& \bibinfo{author}{Bertsch, G.~F.}
\newblock \bibinfo{title}{{R}eal-time, real-space implementation of the linear
  response time-dependent density-functional theory}.
\newblock \emph{\bibinfo{journal}{Phys. Status Solidi B}}
  \textbf{\bibinfo{volume}{243}}, \bibinfo{pages}{1121} (\bibinfo{year}{2006}).

\bibitem{Otobe2008}
\bibinfo{author}{Otobe, T.} \emph{et~al.}
\newblock \bibinfo{title}{{F}irst-principles electron dynamics simulation for
  optical breakdown of dielectrics under an intense laser field}.
\newblock \emph{\bibinfo{journal}{Phys. Rev. B}} \textbf{\bibinfo{volume}{77}},
  \bibinfo{pages}{165104} (\bibinfo{year}{2008}).

\bibitem{Otobe2009}
\bibinfo{author}{Otobe, T.}, \bibinfo{author}{Yabana, K.} \&
  \bibinfo{author}{Iwata, J.-I.}
\newblock \bibinfo{title}{{F}irst-principles calculation of the electron
  dynamics in crystalline {{S}i{O}}2}.
\newblock \emph{\bibinfo{journal}{J. Phys. Condens. Matter}}
  \textbf{\bibinfo{volume}{21}}, \bibinfo{pages}{064224}
  (\bibinfo{year}{2009}).

\bibitem{Otobe2016b}
\bibinfo{author}{Otobe, T.}, \bibinfo{author}{Shinohara, Y.},
  \bibinfo{author}{Sato, S.~A.} \& \bibinfo{author}{Yabana, K.}
\newblock \bibinfo{title}{{F}emtosecond time-resolved dynamical
  {F}ranz-{K}eldysh effect}.
\newblock \emph{\bibinfo{journal}{Phys. Rev. B}} \textbf{\bibinfo{volume}{93}},
  \bibinfo{pages}{045124} (\bibinfo{year}{2016}).

\bibitem{Yabana2012}
\bibinfo{author}{Yabana, K.}, \bibinfo{author}{Sugiyama, T.},
  \bibinfo{author}{Shinohara, Y.}, \bibinfo{author}{Otobe, T.} \&
  \bibinfo{author}{Bertsch, G.~F.}
\newblock \bibinfo{title}{{T}ime-dependent density functional theory for strong
  electromagnetic fields in crystalline solids}.
\newblock \emph{\bibinfo{journal}{Phys. Rev. B}} \textbf{\bibinfo{volume}{85}},
  \bibinfo{pages}{045134} (\bibinfo{year}{2012}).

\bibitem{Shinohara2010}
\bibinfo{author}{Shinohara, Y.} \emph{et~al.}
\newblock \bibinfo{title}{{F}irst-principles description for coherent phonon
  generation in diamond}.
\newblock \emph{\bibinfo{journal}{J. Phys. Condens. Matter}}
  \textbf{\bibinfo{volume}{22}}, \bibinfo{pages}{384212}
  (\bibinfo{year}{2010}).

\bibitem{Shinohara2010a}
\bibinfo{author}{Shinohara, Y.} \emph{et~al.}
\newblock \bibinfo{title}{{C}oherent phonon generation in time-dependent
  density functional theory}.
\newblock \emph{\bibinfo{journal}{Phys. Rev. B}} \textbf{\bibinfo{volume}{82}},
  \bibinfo{pages}{155110} (\bibinfo{year}{2010}).

\bibitem{Shinohara2012}
\bibinfo{author}{Shinohara, Y.} \emph{et~al.}
\newblock \bibinfo{title}{{N}onadiabatic generation of coherent phonons}.
\newblock \emph{\bibinfo{journal}{J. Chem. Phys.}}
  \textbf{\bibinfo{volume}{137}}, \bibinfo{pages}{22A527}
  (\bibinfo{year}{2012}).

\bibitem{Sato2015}
\bibinfo{author}{Sato, S.~A.} \emph{et~al.}
\newblock \bibinfo{title}{{T}ime-dependent density functional theory of
  high-intensity short-pulse laser irradiation on insulators}.
\newblock \emph{\bibinfo{journal}{Phys. Rev. B}} \textbf{\bibinfo{volume}{92}},
  \bibinfo{pages}{205413} (\bibinfo{year}{2015}).

\bibitem{Sato2015a}
\bibinfo{author}{Sato, S.~A.}, \bibinfo{author}{Taniguchi, Y.},
  \bibinfo{author}{Shinohara, Y.} \& \bibinfo{author}{Yabana, K.}
\newblock \bibinfo{title}{{N}onlinear electronic excitations in crystalline
  solids using meta-generalized gradient approximation and hybrid functional in
  time-dependent density functional theory}.
\newblock \emph{\bibinfo{journal}{J. Chem. Phys.}}
  \textbf{\bibinfo{volume}{143}}, \bibinfo{pages}{224116}
  (\bibinfo{year}{2015}).

\bibitem{Ren2013}
\bibinfo{author}{Ren, J.}, \bibinfo{author}{Vukmirovi{\'{c}}, N.} \&
  \bibinfo{author}{Wang, L.-W.}
\newblock \bibinfo{title}{{N}onadiabatic molecular dynamics simulation for
  carrier transport in a pentathiophene butyric acid monolayer}.
\newblock \emph{\bibinfo{journal}{Phys. Rev. B}} \textbf{\bibinfo{volume}{87}},
  \bibinfo{pages}{205117} (\bibinfo{year}{2013}).

\bibitem{Wang2015a}
\bibinfo{author}{Wang, Z.}, \bibinfo{author}{Li, S.-S.} \&
  \bibinfo{author}{Wang, L.-W.}
\newblock \bibinfo{title}{{E}fficient {R}eal-{T}ime {T}ime-{D}ependent
  {D}ensity {F}unctional {T}heory {M}ethod and its {A}pplication to a
  {C}ollision of an {I}on with a 2{D} {M}aterial}.
\newblock \emph{\bibinfo{journal}{Phys. Rev. Lett.}}
  \textbf{\bibinfo{volume}{114}}, \bibinfo{pages}{063004}
  (\bibinfo{year}{2015}).

\bibitem{DiSalvo1976}
\bibinfo{author}{Salvo, F. J.~D.}, \bibinfo{author}{Moncton, D.~E.} \&
  \bibinfo{author}{Waszczak, J.~V.}
\newblock \bibinfo{title}{{E}lectronic properties and superlattice formation in
  the semimetal {T}i{S}e$_2$}.
\newblock \emph{\bibinfo{journal}{Phys. Rev. B}} \textbf{\bibinfo{volume}{14}},
  \bibinfo{pages}{4321} (\bibinfo{year}{1976}).

\bibitem{Bianco2015}
\bibinfo{author}{Bianco, R.}, \bibinfo{author}{Calandra, M.} \&
  \bibinfo{author}{Mauri, F.}
\newblock \bibinfo{title}{{E}lectronic and vibrational properties {of
  {T}i{S}e}$_2$ in the charge-density-wave phase from first principles}.
\newblock \emph{\bibinfo{journal}{Phys. Rev. B}} \textbf{\bibinfo{volume}{92}},
  \bibinfo{pages}{094107} (\bibinfo{year}{2015}).

\bibitem{Sharma2011}
\bibinfo{author}{Sharma, S.}, \bibinfo{author}{Dewhurst, J.~K.},
  \bibinfo{author}{Sanna, A.} \& \bibinfo{author}{Gross, E. K.~U.}
\newblock \bibinfo{title}{{B}ootstrap {A}pproximation for the
  {E}xchange-{C}orrelation {K}ernel of {T}ime-{D}ependent
  {D}ensity-{F}unctional {T}heory}.
\newblock \emph{\bibinfo{journal}{Phys. Rev. Lett.}}
  \textbf{\bibinfo{volume}{107}}, \bibinfo{pages}{186401}
  (\bibinfo{year}{2011}).

\bibitem{Sottile2003}
\bibinfo{author}{Sottile, F.}, \bibinfo{author}{Olevano, V.} \&
  \bibinfo{author}{Reining, L.}
\newblock \bibinfo{title}{{Parameter-Free Calculation of Response Functions in
  Time-Dependent Density-Functional Theory}}.
\newblock \emph{\bibinfo{journal}{Phys. Rev. Lett.}}
  \textbf{\bibinfo{volume}{91}}, \bibinfo{pages}{056402}
  (\bibinfo{year}{2003}).

\bibitem{Marini2003}
\bibinfo{author}{Marini, A.}, \bibinfo{author}{{Del Sole}, R.} \&
  \bibinfo{author}{Rubio, A.}
\newblock \bibinfo{title}{{Bound Excitons in Time-Dependent Density-Functional
  Theory: Optical and Energy-Loss Spectra}}.
\newblock \emph{\bibinfo{journal}{Phys. Rev. Lett.}}
  \textbf{\bibinfo{volume}{91}}, \bibinfo{pages}{256402}
  (\bibinfo{year}{2003}).

\bibitem{Weber2011}
\bibinfo{author}{Weber, F.} \emph{et~al.}
\newblock \bibinfo{title}{{E}lectron-{P}honon {C}oupling and the {S}oft
  {P}honon {M}ode {in {T}i{S}e}$_2$}.
\newblock \emph{\bibinfo{journal}{Phys. Rev. Lett.}}
  \textbf{\bibinfo{volume}{107}}, \bibinfo{pages}{266401}
  (\bibinfo{year}{2011}).

\bibitem{Singh2017a}
\bibinfo{author}{Singh, B.}, \bibinfo{author}{Hsu, C.-H.},
  \bibinfo{author}{Tsai, W.-F.}, \bibinfo{author}{Pereira, V.~M.} \&
  \bibinfo{author}{Lin, H.}
\newblock \bibinfo{title}{Stable charge density wave phase in a
  1{T}{\textendash}{TiSe}$_2$ monolayer}.
\newblock \emph{\bibinfo{journal}{Phys. Rev. B}} \textbf{\bibinfo{volume}{95}},
  \bibinfo{pages}{245136} (\bibinfo{year}{2017}).

\bibitem{Hellgren2017}
\bibinfo{author}{Hellgren, M.} \emph{et~al.}
\newblock \bibinfo{title}{Critical role of the exchange interaction for the
  electronic structure and charge-density-wave formation in {{T}i{S}e}$_2$}.
\newblock \emph{\bibinfo{journal}{Phys. Rev. Lett.}}
  \textbf{\bibinfo{volume}{119}}, \bibinfo{pages}{176401}
  (\bibinfo{year}{2017}).

\bibitem{Snow2003}
\bibinfo{author}{Snow, C.~S.}, \bibinfo{author}{Karpus, J.~F.},
  \bibinfo{author}{Cooper, S.~L.}, \bibinfo{author}{Kidd, T.~E.} \&
  \bibinfo{author}{Chiang, T.-C.}
\newblock \bibinfo{title}{{Q}uantum {M}elting of the {C}harge-{D}ensity-{W}ave
  {S}tate in 1{T}-{{T}i{S}e}$_2$}.
\newblock \emph{\bibinfo{journal}{Phys. Rev. Lett.}}
  \textbf{\bibinfo{volume}{91}}, \bibinfo{pages}{136402}
  (\bibinfo{year}{2003}).

\bibitem{Wang2018g}
\bibinfo{author}{Wang, H.} \emph{et~al.}
\newblock \bibinfo{title}{Large-area atomic layers of the charge-density-wave
  conductor {TiSe}2}.
\newblock \emph{\bibinfo{journal}{Adv. Mater.}} \textbf{\bibinfo{volume}{30}},
  \bibinfo{pages}{1704382} (\bibinfo{year}{2018}).

\bibitem{Floss2017}
\bibinfo{author}{Floss, I.} \emph{et~al.}
\newblock \bibinfo{title}{Ab initio multiscale simulation of high-order
  harmonic generation in solids}.
\newblock \emph{\bibinfo{journal}{Phys. Rev. A}} \textbf{\bibinfo{volume}{97}},
  \bibinfo{pages}{011401} (\bibinfo{year}{2018}).

\bibitem{Yamada2018}
\bibinfo{author}{Yamada, S.}, \bibinfo{author}{Noda, M.},
  \bibinfo{author}{Nobusada, K.} \& \bibinfo{author}{Yabana, K.}
\newblock \bibinfo{title}{{T}ime-dependent density functional theory for
  interaction of ultrashort light pulse with thin materials}.
\newblock \emph{\bibinfo{journal}{Phys. Rev. B}} \textbf{\bibinfo{volume}{98}},
  \bibinfo{pages}{245147} (\bibinfo{year}{2018}).

\bibitem{Floss2019}
\bibinfo{author}{Floss, I.}, \bibinfo{author}{Lemell, C.},
  \bibinfo{author}{Yabana, K.} \& \bibinfo{author}{Burgd\"{o}rfer, J.}
\newblock \bibinfo{title}{Incorporating decoherence into solid-state
  time-dependent density functional theory}.
\newblock \emph{\bibinfo{journal}{Phys. Rev. B}} \textbf{\bibinfo{volume}{99}}
  (\bibinfo{year}{2019}).

\bibitem{Tully1990}
\bibinfo{author}{Tully, J.~C.}
\newblock \bibinfo{title}{{M}olecular dynamics with electronic transitions}.
\newblock \emph{\bibinfo{journal}{J. Chem. Phys.}}
  \textbf{\bibinfo{volume}{93}}, \bibinfo{pages}{1061} (\bibinfo{year}{1990}).

\bibitem{Gossel2018}
\bibinfo{author}{Gossel, G.~H.}, \bibinfo{author}{Agostini, F.} \&
  \bibinfo{author}{Maitra, N.~T.}
\newblock \bibinfo{title}{{C}oupled-{T}rajectory {M}ixed {Q}uantum-{C}lassical
  {A}lgorithm: {A} {D}econstruction}.
\newblock \emph{\bibinfo{journal}{J. Chem. Theory. Comput.}}
  \textbf{\bibinfo{volume}{14}}, \bibinfo{pages}{4513} (\bibinfo{year}{2018}).

\bibitem{Zhang2018CDWTaS2}
\bibinfo{author}{Zhang, J.} \emph{et~al.}
\newblock \bibinfo{title}{Photoexcitation induced quantum dynamics of charge
  density wave and emergence of a collective mode in 1{T}-{TaS}2}.
\newblock \emph{\bibinfo{journal}{Nano Lett.}} \textbf{\bibinfo{volume}{19}},
  \bibinfo{pages}{6027--6034} (\bibinfo{year}{2019}).

\bibitem{Lian2019Silicon}
\bibinfo{author}{Lian, C.}, \bibinfo{author}{Zhang, S.~B.} \&
  \bibinfo{author}{Meng, S.}
\newblock \bibinfo{title}{{U}ltrafast carrier relaxation and its {P}auli drag
  in photo-enhanced melting of solids} (\bibinfo{year}{2019}).
\newblock \urlprefix\url{https://arxiv.org/abs/1901.00609}.
\newblock \eprint{1901.00609}.

\bibitem{Lian2016NTM}
\bibinfo{author}{Lian, C.}, \bibinfo{author}{Zhang, S.~B.} \&
  \bibinfo{author}{Meng, S.}
\newblock \bibinfo{title}{Ab initio evidence for nonthermal characteristics in
  ultrafast laser melting}.
\newblock \emph{\bibinfo{journal}{Phys. Rev. B}} \textbf{\bibinfo{volume}{94}},
  \bibinfo{pages}{184310} (\bibinfo{year}{2016}).

\bibitem{Meng2008a}
\bibinfo{author}{Meng, S.} \& \bibinfo{author}{Kaxiras, E.}
\newblock \bibinfo{title}{{R}eal-time, local basis-set implementation of
  time-dependent density functional theory for excited state dynamics
  simulations}.
\newblock \emph{\bibinfo{journal}{J. Chem. Phys.}}
  \textbf{\bibinfo{volume}{129}}, \bibinfo{pages}{054110}
  (\bibinfo{year}{2008}).

\bibitem{Lian2018MultiK}
\bibinfo{author}{Lian, C.}, \bibinfo{author}{Hu, S.-Q.}, \bibinfo{author}{Guan,
  M.-X.} \& \bibinfo{author}{Meng, S.}
\newblock \bibinfo{title}{{M}omentum-resolved {{T}{D}{D}{F}{T}} algorithm in
  atomic basis for real time tracking of electronic excitation}.
\newblock \emph{\bibinfo{journal}{J. Chem. Phys.}}
  \textbf{\bibinfo{volume}{149}}, \bibinfo{pages}{154104}
  (\bibinfo{year}{2018}).

\bibitem{Lian2018AdvTheo}
\bibinfo{author}{Lian, C.}, \bibinfo{author}{Guan, M.}, \bibinfo{author}{Hu,
  S.}, \bibinfo{author}{Zhang, J.} \& \bibinfo{author}{Meng, S.}
\newblock \bibinfo{title}{{P}hotoexcitation in {S}olids: {F}irst-{P}rinciples
  {Q}uantum {S}imulations by {R}eal-{T}ime {{T}{D}{D}{F}{T}}}.
\newblock \emph{\bibinfo{journal}{Advanced Theory and Simulations}}
  \textbf{\bibinfo{volume}{1}}, \bibinfo{pages}{1800055}
  (\bibinfo{year}{2018}).
\newblock
  \urlprefix\url{https://onlinelibrary.wiley.com/doi/full/10.1002/adts.201800055}.

\bibitem{Giannozzi2009}
\bibinfo{author}{Giannozzi, P.} \emph{et~al.}
\newblock \bibinfo{title}{{{Q}{U}{A}{N}{T}{U}{M}} {{E}{S}{P}{R}{E}{S}{S}{O}}: a
  modular and open-source software project for quantum simulations of
  materials}.
\newblock \emph{\bibinfo{journal}{J. Phys. Condens. Matter}}
  \textbf{\bibinfo{volume}{21}}, \bibinfo{pages}{395502}
  (\bibinfo{year}{2009}).

\bibitem{Blochl1994}
\bibinfo{author}{Blöchl, P.~E.}
\newblock \bibinfo{title}{{P}rojector augmented-wave method}.
\newblock \emph{\bibinfo{journal}{Phys. Rev. B}} \textbf{\bibinfo{volume}{50}},
  \bibinfo{pages}{17953} (\bibinfo{year}{1994}).

\bibitem{Perdew1996}
\bibinfo{author}{Perdew, J.~P.}, \bibinfo{author}{Burke, K.} \&
  \bibinfo{author}{Ernzerhof, M.}
\newblock \bibinfo{title}{{G}eneralized {G}radient {A}pproximation {M}ade
  {S}imple}.
\newblock \emph{\bibinfo{journal}{Phys. Rev. Lett.}}
  \textbf{\bibinfo{volume}{77}}, \bibinfo{pages}{3865} (\bibinfo{year}{1996}).

\bibitem{DalCorso2014}
\bibinfo{author}{Corso, A.~D.}
\newblock \bibinfo{title}{{P}seudopotentials periodic table: {F}rom {H} to
  {P}u}.
\newblock \emph{\bibinfo{journal}{Computational Materials Science}}
  \textbf{\bibinfo{volume}{95}}, \bibinfo{pages}{337} (\bibinfo{year}{2014}).

\bibitem{Monkhorst1976}
\bibinfo{author}{Monkhorst, H.~J.} \& \bibinfo{author}{Pack, J.~D.}
\newblock \bibinfo{title}{{S}pecial points for {B}rillouin-zone integrations}.
\newblock \emph{\bibinfo{journal}{Phys. Rev. B}} \textbf{\bibinfo{volume}{13}},
  \bibinfo{pages}{5188} (\bibinfo{year}{1976}).

\bibitem{Ku2010}
\bibinfo{author}{Ku, W.}, \bibinfo{author}{Berlijn, T.} \&
  \bibinfo{author}{Lee, C.-C.}
\newblock \bibinfo{title}{{U}nfolding {F}irst-{P}rinciples {B}and
  {S}tructures}.
\newblock \emph{\bibinfo{journal}{Phys. Rev. Lett.}}
  \textbf{\bibinfo{volume}{104}}, \bibinfo{pages}{216401}
  (\bibinfo{year}{2010}).

\bibitem{Popescu2012}
\bibinfo{author}{Popescu, V.} \& \bibinfo{author}{Zunger, A.}
\newblock \bibinfo{title}{{{E}xtracting {E} versus k} effective band structure
  from supercell calculations on alloys and impurities}.
\newblock \emph{\bibinfo{journal}{Phys. Rev. B}} \textbf{\bibinfo{volume}{85}},
  \bibinfo{pages}{085201} (\bibinfo{year}{2012}).

\bibitem{Medeiros2014}
\bibinfo{author}{Medeiros, P. V.~C.}, \bibinfo{author}{Stafstr\"{o}m, S.} \&
  \bibinfo{author}{Bj\"{o}rk, J.}
\newblock \bibinfo{title}{{E}ffects of extrinsic and intrinsic perturbations on
  the electronic structure of graphene: {R}etaining an effective primitive cell
  band structure by band unfolding}.
\newblock \emph{\bibinfo{journal}{Phys. Rev. B}} \textbf{\bibinfo{volume}{89}},
  \bibinfo{pages}{041407} (\bibinfo{year}{2014}).

\bibitem{Medeiros2015}
\bibinfo{author}{Medeiros, P. V.~C.}, \bibinfo{author}{Tsirkin, S.~S.},
  \bibinfo{author}{Stafstr\"{o}m, S.} \& \bibinfo{author}{Bj\"{o}rk, J.}
\newblock \bibinfo{title}{{U}nfolding spinor wave functions and expectation
  values of general operators: {I}ntroducing the unfolding-density operator}.
\newblock \emph{\bibinfo{journal}{Phys. Rev. B}} \textbf{\bibinfo{volume}{91}},
  \bibinfo{pages}{041116} (\bibinfo{year}{2015}).

\bibitem{Lian2017SiUnfold}
\bibinfo{author}{Lian, C.} \& \bibinfo{author}{Meng, S.}
\newblock \bibinfo{title}{{D}irac cone pairs in silicene induced by interface
  {S}i-{A}g hybridization: {A} first-principles effective band study}.
\newblock \emph{\bibinfo{journal}{Phys. Rev. B}} \textbf{\bibinfo{volume}{95}},
  \bibinfo{pages}{245409} (\bibinfo{year}{2017}).

\bibitem{Marini2009}
\bibinfo{author}{Marini, A.}, \bibinfo{author}{Hogan, C.},
  \bibinfo{author}{Gr{\"{u}}ning, M.} \& \bibinfo{author}{Varsano, D.}
\newblock \bibinfo{title}{{yambo: An ab initio tool for excited state
  calculations}}.
\newblock \emph{\bibinfo{journal}{Comput. Phys. Commun.}}
  \textbf{\bibinfo{volume}{180}}, \bibinfo{pages}{1392--1403}
  (\bibinfo{year}{2009}).

\bibitem{Nose1984}
\bibinfo{author}{Nos{\'{e}}, S.}
\newblock \bibinfo{title}{{A unified formulation of the constant temperature
  molecular dynamics methods}}.
\newblock \emph{\bibinfo{journal}{J. Chem. Phys.}}
  \textbf{\bibinfo{volume}{81}}, \bibinfo{pages}{511--519}
  (\bibinfo{year}{1984}).

\bibitem{Hoover1985}
\bibinfo{author}{Hoover, W.~G.}
\newblock \bibinfo{title}{{Canonical dynamics: Equilibrium phase-space
  distributions}}.
\newblock \emph{\bibinfo{journal}{Phys. Rev. A}} \textbf{\bibinfo{volume}{31}},
  \bibinfo{pages}{1695--1697} (\bibinfo{year}{1985}).

\bibitem{Berendsen1984}
\bibinfo{author}{Berendsen, H. J.~C.}, \bibinfo{author}{Postma, J. P.~M.},
  \bibinfo{author}{van Gunsteren, W.~F.}, \bibinfo{author}{DiNola, A.} \&
  \bibinfo{author}{Haak, J.~R.}
\newblock \bibinfo{title}{{Molecular dynamics with coupling to an external
  bath}}.
\newblock \emph{\bibinfo{journal}{J. Chem. Phys.}}
  \textbf{\bibinfo{volume}{81}}, \bibinfo{pages}{3684--3690}
  (\bibinfo{year}{1984}).

\end{thebibliography}

\beginsupplement
\widetext
\section{Algorithms of TDDFT in Planewave basis}

\subsection{Planewave and Adiabatic Basis}
Following our previous scenario in the Time dependent \textit{ab initio} package (TDAP)~\cite{Meng2008a,Lian2018MultiK,Lian2018AdvTheo}, we implement the TDDFT algorithm in PW basis. The time-dependent Kohn-Sham (TDKS) equation at time $t$ in PW basis $\{\mathbf{G}\}$ reads~\cite{Runge1984}:
\begin{equation}
\label{eq.TDKS}
i\hbar\frac{\partial \psi_{\gamma \mathbf{k}}(\mathbf{G},t)}{\partial t} = \mathcal{H}_{\mathbf{k}}(t) \psi_{\gamma \mathbf{k}}(\mathbf{G},t)
\end{equation}
where $\psi_{\gamma \mathbf{k}}(\mathbf{G},t)$ is TDKS orbital, $\gamma$ denotes the band index, $\mathbf{k}$ is the reciprocal momentum index.
$\mathcal{H}_{\mathbf{k}}(t)$ is the Hamiltonian expanded with plane-wave basis, with matrix element 
\begin{equation}
\begin{split}
\mathcal{H}_{\mathbf{k}}(\mathbf{G},\mathbf{G'},t) =& T_\mathbf{k}(\mathbf{G},\mathbf{G'},t) + V(\mathbf{G},\mathbf{G'},t) \\
=& \frac{\hbar^2}{2m}|\mathbf{k}+\mathbf{G}+\mathbf{A}(t)|^2 \delta_{\mathbf{G},\mathbf{G'}} + V(\mathbf{G},\mathbf{G'},t)
\end{split}
\end{equation} 
where $T_\mathbf{k}(\mathbf{G},\mathbf{G'}) = \frac{\hbar^2}{2m}|\mathbf{k}+\mathbf{G}+\mathbf{A}(t)|^2 \delta_{\mathbf{G},\mathbf{G'}}$ is the kinetic term, $\mathbf{A}$ is the velocity gauge potential~\cite{Bertsch2000,Yabana2006}:
\begin{equation}
\mathbf{A}(t) = -\int_0^t \mathbf{E}(t') dt'
\end{equation}
where $\mathbf{E}$ is electric field. $V(\mathbf{G},\mathbf{G'})$ is the potential term calculated within the corresponding module \textsf{Quantum Espresso}, including such as ion-electron potential, Hartree potential and exchange-correlation potential. 

There are $N_b$ independent TDKS equations for each $\mathbf{k}$ index, and thus $N_bN_\mathbf{k}$ equations in total, where $N_b$ is the number of bands, typically, $N_\mathbf{k}$ is the number of $\mathbf{k}$ points. Typically, the orders of $N_\mathbf{k}$ and $N_b$ are both $\sim 10^2$. The number of $\{\mathbf{G}\}$, $N_G$, is usually $\sim 10^4$. Thus, if the number of time steps is $N_t$, the complexity of solving Eq.~(\ref{eq.TDKS}) is $O(N_tN_bN_\mathbf{k}N^2_G)$.

We express the TDKS orbitals $\psi_{\gamma \mathbf{k}}(\mathbf{G},t)$ using adiabatic basis $\{\phi_{\gamma \mathbf{k}}(\mathbf{G},t_1)\}$
\begin{equation}
\label{eq:adexpand}
\ket{\psi_{\gamma \mathbf{k}}(\mathbf{G},t)} = \sum_i c_{i\gamma, \mathbf{k}}(t) \ket{\phi_{i \mathbf{k}}(\mathbf{G},t_1)}
\end{equation}
where $c_{i\gamma, \mathbf{k}}(t)$ is the TD coefficients and the adiabatic basis $\{\phi_{i \mathbf{k}}\}(t_1)$ are solved by diagonalizing the Hamiltonian
\begin{equation}
\label{eq:diagon}
\mathcal{H}_{\mathbf{k}}(\mathbf{G},t_1) \ket{\phi_{i \mathbf{k}}(\mathbf{G},t_1)} = \epsilon_{i\mathbf{k}}(t_1) \ket{\phi_{i \mathbf{k}}(\mathbf{G},t_1)},
\end{equation}
where the $\epsilon_{i\mathbf{k}}$ is the eigenvalue.
The initial condition is chosen as 
\begin{equation}
\begin{split}
\ket{\psi_{i \mathbf{k}}(\mathbf{G},t=0)} :=& \ket{\phi_{i \mathbf{k}}(\mathbf{G},t_1=0)}, \\
c_{i\gamma,\mathbf{k}}(t=0) :=& \delta_{i\gamma},\\
\end{split}
\end{equation}
where $\delta_{i\gamma}$ is Kronecker delta.
The TDKS equations are interpreted with the evolution of coefficient matrix 
\begin{equation}
\label{eq:coeffTDKS}
{H}_\mathbf{k}(t){C}_\mathbf{k}(t) = i\hbar\frac{\partial }{\partial t}{C}_\mathbf{k}(t), 
\end{equation}
where 
\begin{equation}
{C}_\mathbf{k}(t) = 
\left(\begin{array}{cccc}
c_{11,\mathbf{k}}(t) &c_{12,\mathbf{k}}(t) & \cdots &c_{1N_b,\mathbf{k}}(t) \\
c_{21,\mathbf{k}}(t) &c_{12,\mathbf{k}}(t) & \cdots &c_{1N_b,\mathbf{k}}(t) \\
\vdots &\vdots & \cdots &\vdots \\
c_{N_b1,\mathbf{k}}(t) &c_{N_b2,\mathbf{k}}(t) & \cdots &c_{N_bN_b,\mathbf{k}}(t) \\
\end{array}\right),
\end{equation}
\begin{equation}
{H}_\mathbf{k}(t) = 
\left(\begin{array}{cccc}
h_{11,\mathbf{k}}(t) &h_{12,\mathbf{k}}(t) & \cdots &h_{1N_b,\mathbf{k}}(t) \\
h_{21,\mathbf{k}}(t) &h_{12,\mathbf{k}}(t) & \cdots &h_{1N_b,\mathbf{k}}(t) \\
\vdots &\vdots & \cdots &\vdots \\
h_{N_b1,\mathbf{k}}(t) &h_{N_b2,\mathbf{k}}(t) & \cdots &h_{N_bN_b,\mathbf{k}}(t) \\
\end{array}\right),
\end{equation}
and $h_{ij,\mathbf{k}} = \braket{\phi_{i\mathbf{k}}(t_1)|\mathcal{H}_\mathbf{k}(t)|\phi_{j\mathbf{k}}(t_1)}$.  To distinguish, we use $\mathcal{A}$ and $A$ to represent the matrix $A$ (e.g. Hamiltonian) in the PW basis and the adiabatic basis, respectively. 

Note that, no approximation has yet been introduced in the derivation. The complexity of solving the Eq.~\ref{eq:diagon} is the same as solving the Eq.~\ref{eq.TDKS}, while it is less time consuming to solve Eq.~\ref{eq:coeffTDKS} noticing that the dimension of ${C}_\mathbf{k}(t)$ and ${H}_\mathbf{k}(t)$ is $N_b\times N_b$ and $N_b \ll N_{G}$. The computation will be accelerated if the time of solving of Eq.~\ref{eq:diagon} is minimized.

\subsection{Efficient Evolution in Adiabatic Basis}
An efficient scheme is raised by Wang \textit{et. al.}~\cite{Wang2015a}: Hamiltonian on adiabatic basis $h_{ij,\mathbf{k}}(t) = \braket{\phi_{i\mathbf{k}}(t_1)|\mathcal{H}_\mathbf{k}(t)|\phi_{j\mathbf{k}}(t_1)}$ changes approximately linearly within $[t_1, t_2]$ with $t_2 = t_1 + \Delta t$~fs and $\Delta t \sim 0.2$~fs~\cite{Ren2013}: 
\begin{equation}
\label{eq.interph}
H_\mathbf{k}(t) = H_\mathbf{k}(t_1) + \frac{t-t_1}{t_2-t_1}[H_\mathbf{k}(t_2) - H_\mathbf{k}(t_1)],
\end{equation}
The matrix element $h_{ij,\mathbf{k}}$ is evaluated as 
\begin{equation}
\label{eq.ht1}
h_{ij,\mathbf{k}}(t_1) = \braket{\phi_{i}(t_1)|\mathcal{H}_\mathbf{k}(t_1)|\phi_{j}(t_1)} = \delta_{ij} \epsilon_{i,\mathbf{k}}(t_1),
\end{equation}
and 
\begin{equation}
\label{eq.ht2}
\begin{split}
H_{ij,\mathbf{k}}(t_2) =& \braket{\phi_{i,\mathbf{k}}(t_1)|\mathcal{H}_\mathbf{k}(t_2)|\phi_{j,\mathbf{k}}(t_1)} \\
=& \braket{\phi_{i,\mathbf{k}}(t_1)|\sum_l \phi_{l,\mathbf{k}}(t_2)}\braket{\phi_{l,\mathbf{k}}(t_2)|\mathcal{H}_\mathbf{k}(t_2)|\sum_m \phi_{m,\mathbf{k}}(t_2)} \braket{\phi_{m,\mathbf{k}}(t_2)|\phi_{j,\mathbf{k}}(t_1)} \\
=& \sum_l\sum_{m} a_{il,\mathbf{k}}(t_1,t_2)a^*_{mj,\mathbf{k}}(t_1,t_2) \braket{\phi_{l,\mathbf{k}}(t_2)|\mathcal{H}_\mathbf{k}(t_2)|\phi_{m,\mathbf{k}}(t_2)}  \\ 
=& \sum_l\sum_{m} a_{il,\mathbf{k}}(t_1,t_2) \delta_{lm} \epsilon_{l\mathbf{k}}(t_2) a_{il,\mathbf{k}}(t_1,t_2) a^*_{mj,\mathbf{k}}(t_1,t_2) \\
=& \sum_l a_{il,\mathbf{k}}(t_1,t_2) a^*_{lj,\mathbf{k}}(t_1,t_2) \epsilon_{l\mathbf{k}}(t_2),
\end{split}
\end{equation} 
where 
\begin{equation}
\label{innerProd}
a_{il,\mathbf{k}}(t_1,t_2) = \braket{\phi_{i,\mathbf{k}}(t_1)|\phi_{l,\mathbf{k}}(t_2)}
\end{equation}
$\ket{\phi_{l,\mathbf{k}}(t_2)}$ and $\epsilon_{l\mathbf{k}}(t_2)$  are the adiabatic basis and the eigenvalue at time $t_2$, which are solved in a self-consistent process, see detail at Sec.~\ref{SCF} .

Propagator operator $U_\mathbf{k}(t_2, t_1)$ is calculated with the knowledge of $H_\mathbf{k}(t)$ in the Crank-Nicholson scheme
\begin{equation}
\label{eq.propagator}
U_\mathbf{k}(t_2, t_1) = \prod_{s=0}^{N_t} \frac{\exp[-i\hbar H(t_1 + sdt)dt/2]}{\exp[i\hbar H(t_1 + sdt)dt/2]}= \prod_{s=0}^{N_t} \frac{1 - {i\hbar } H_\mathbf{k}(t_1 + sdt)dt /2}{1 + {i \hbar }  H_\mathbf{k}(t_1 + sdt)dt /2},
\end{equation}
where $dt = \Delta t/N_t$ is the integration time step. Since $dtH \ll 1$ is needed to satisfy the condition of $\exp(-i\hbar dtH) = 1 - {i\hbar } H dt$,  we choose $dt\sim 0.1$~attosecond, considering H is about $10^2$~eV.  The coefficients $c_{i\gamma,\mathbf{k}}$ are thus propagated as
\begin{equation}
\label{eq.propagating}
{C}_\mathbf{k}(t_2) = U_\mathbf{k}(t_2, t_1) {C}_\mathbf{k}(t_1).
\end{equation}
Thus, we finish the evolution of TDKS orbital from $t_1$ to $t_2$ as
\begin{equation}
\label{eq:suddenchange}
\begin{split}
\ket{\psi_{\gamma \mathbf{k}}(t_2)} &= \sum_i c_{i\gamma,\mathbf{k}}(t_2) \ket{\phi_{i,\mathbf{k}}(t_1)} \\
&= \sum_i c_{i\gamma,\mathbf{k}}(t_2) \sum_l \ket{\phi_{l,\mathbf{k}}(t_2)}\braket{\phi_{l,\mathbf{k}}(t_2)|\phi_{i,\mathbf{k}}(t_1)}\\
&= \sum_l \sum_i a^*_{il,\mathbf{k}}(t_1,t_2) c_{i\gamma,\mathbf{k}}(t_2)\ket{\phi_{l,\mathbf{k}}(t_2)} \\ 
&= \sum_l c'_{l\gamma,\mathbf{k}}(t_2) \ket{\phi_{l,\mathbf{k}}(t_2)}
\end{split}
\end{equation}
where 
\begin{equation}
c'_{l\gamma,\mathbf{k}}(t_2) = \sum_i a^*_{il,\mathbf{k}}(t_1,t_2) c_{i\gamma,\mathbf{k}}(t_2).
\end{equation}
Charge density $\rho(t_2)$ can be calculated with $c'_{l\gamma,\mathbf{k}}(t_2)$ and $\ket{\phi_{l,\mathbf{k}}(t_2)}$ as
\begin{equation}
\label{eq:buildDM}
\begin{split}
\rho(\mathbf{G},t_2) &= \sum_\mathbf{k} \sum_{\gamma} \left| \psi_{\gamma,\mathbf{k}}(\mathbf{G}, t_2) \right|^2 \\
&= \sum_\mathbf{k}\sum_{\gamma} \sum_i |c'_{i\gamma,\mathbf{k}}(t_2)|^2 \left|\phi_{i,\mathbf{k}}(\mathbf{G},t_2)\right|^2 \\
&= \sum_\mathbf{k}\sum_i \left[\sum_{\gamma} |c'_{i\gamma,\mathbf{k}}(t_2)|^2 \right] \left|\phi_{i\mathbf{k}}(\mathbf{G},t_2)\right|^2 \\
&= \sum_\mathbf{k}\sum_i q_{i\mathbf{k}}(t_2) \left|\phi_{i\mathbf{k}}(\mathbf{G})\right|^2,
\end{split}
\end{equation}
where
\begin{equation}
\label{eq:pop}
q_{i\mathbf{k}}(t_2) = \sum_{\gamma} |c'_{i\gamma,\mathbf{k}}(t_2)|^2 
\end{equation}
is the population of the adiabatic states.

\subsection{Computational Flowchart \label{SCF}}
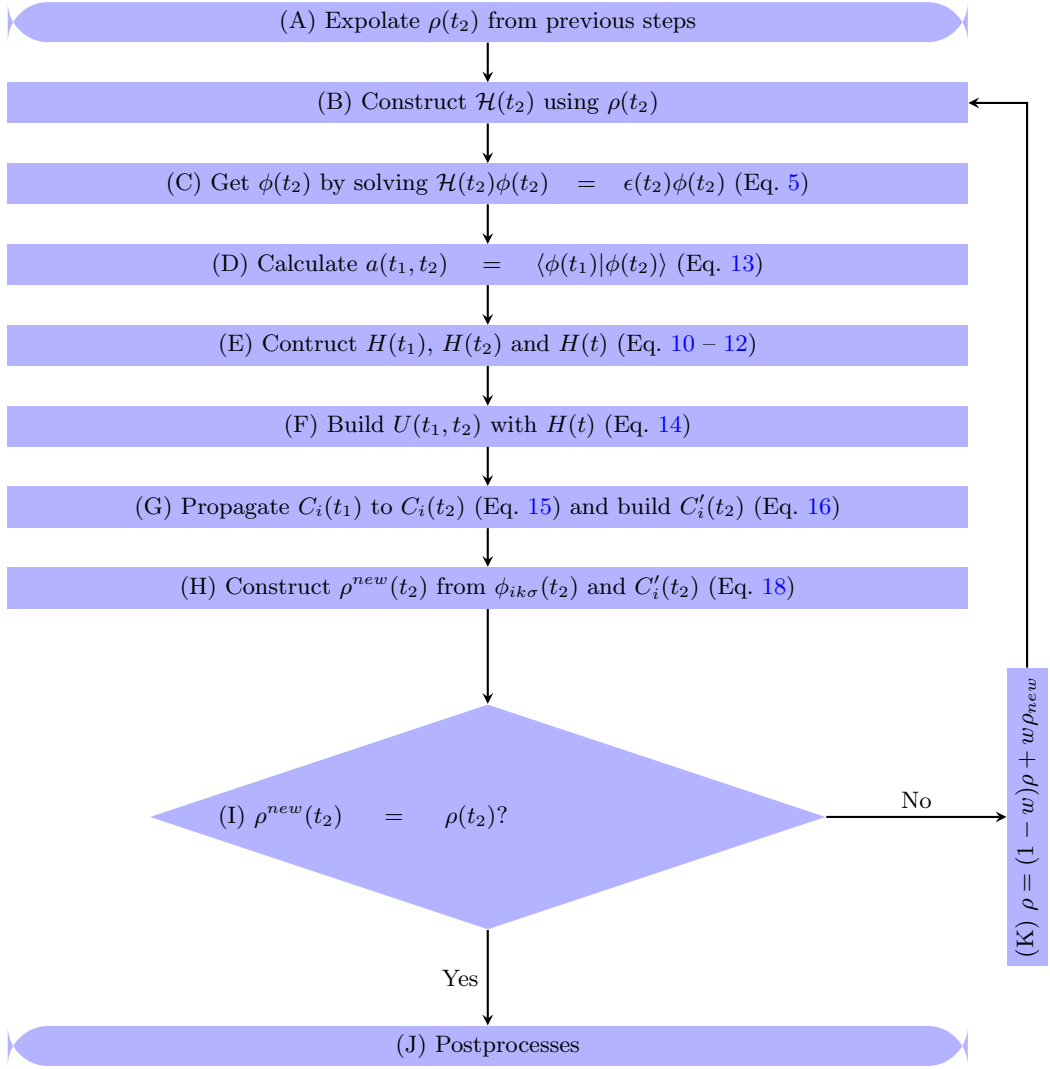
\begin{figure}
	\begin{tikzpicture}[node distance=0.06\textwidth]
	\tikzstyle{startstop} = [rectangle, text centered, rounded corners=0.03\textwidth, fill=blue!30]
	\tikzstyle{basic} = [rectangle, text centered, fill=blue!30]
	\tikzstyle{decision} = [diamond, aspect=3, fill=blue!30]
	\tikzstyle{arrow} = [thick,->,>=stealth]
	\node (start) [left, startstop, text width=0.7\textwidth] {(A) Expolate $\rho(t_2)$ from previous steps};
	\node (in1) [left, basic, below of = start, , text width=0.7\textwidth] {(B) Construct $\mathcal{H}(t_2)$ using $\rho(t_2)$};
	\node (in2) [left, basic, below of = in1, text width=0.7\textwidth] {(C) Get $\phi(t_2)$ by solving $\mathcal{H}(t_2) \phi(t_2) = \epsilon(t_2) \phi(t_2)$ (Eq.~\ref{eq:diagon})};
	\node (in21) [left, basic, below of = in2, text width=0.7\textwidth] {(D) Calculate $a(t_1,t_2) =\braket{\phi(t_1)|\phi(t_2)}$ (Eq.~\ref{innerProd})};
	\node (in22) [left, basic, below of = in21, text width=0.7\textwidth] {(E) Contruct $H(t_1)$, $H(t_2)$ and $H(t)$ (Eq.~\ref{eq.interph} -- \ref{eq.ht2})};
	\node (in23) [left, basic, below of = in22, text width=0.7\textwidth] {(F) Build $U(t_1,t_2)$ with $H(t)$ (Eq.~\ref{eq.propagator})};
	\node (in24) [left, basic, below of = in23, text width=0.7\textwidth] {(G) Propagate $C_i(t_1)$ to $C_i(t_2)$ (Eq.~\ref{eq.propagating}) and build $C'_i(t_2)$ (Eq.~\ref{eq:suddenchange})};
	\node (in3) [left, basic, below of = in24, text width=0.7\textwidth] {(H) Construct $\rho^{new}(t_2)$ from $\phi_{i k \sigma}(t_2)$ and $C'_i(t_2)$ (Eq.~\ref{eq:buildDM}) };
	\node (dec1) [decision, below of = in3, node distance=0.17\textwidth, text width=0.4\textwidth] {(I) $\rho^{new}(t_2) = \rho(t_2)$?};
	\node (pro2a) [startstop, below of=dec1, node distance=0.17\textwidth, text width=0.7\textwidth] {(J) Postprocesses}; 
	\node (pro2b) [basic, right of = dec1, rotate=90,  node distance=0.4\textwidth] {(K) $\rho = (1-w)\rho + w\rho_{new}$};
	\draw [arrow](start) -- (in1);
	\draw [arrow](in1) -- (in2);
	\draw [arrow](in2) -- (in21);
	\draw [arrow](in21) -- (in22) ;
	\draw [arrow](in22) -- (in23) ;
	\draw [arrow](in23) -- (in24) ;
	\draw [arrow](in24) -- (in3);
	\draw [arrow](in3) -- (dec1);
	\draw [arrow](dec1) -- node[anchor=east] {Yes} (pro2a);
	\draw [arrow](dec1) -- node[anchor=south] {No} (pro2b);
	\draw [arrow](pro2b) |- (in1);
	\end{tikzpicture}
	\caption{Flowchart of TDDFT algorithm.}
	\label{fig:flowchart-tdpw}
\end{figure}

We organized the equations in a flowchart, as shown in Fig.~\ref{fig:flowchart-tdpw}. Step (A) (B) (C) (H) (I) (K) are computed with the \textsf{Quantum Espresso} modules. As mentioned above, the algorithm requires foreknowledge of $\rho(\mathbf{G},t_2)$ in the calculation of Hamiltonian $H_\mathbf{k}(t)$. It is satisfied with a self-consistent process:  
\begin{enumerate}
	\item An initial guess of $\rho(\mathbf{G},t_2)$ is built using extrapolation from previous step. 
	\item Hamiltonian $H_\mathbf{k}(t)$ and propagator $U_\mathbf{k}(t_2,t_1)$ are built from $\rho(\mathbf{G},t_2)$ using Eq.~\ref{eq.interph} and Eq.~\ref{eq.propagator}, respectively.
	\item An new $\rho(\mathbf{G},t_2)$ is calculated with new population propagated from Eq.~\ref{eq.propagating}.
	\item The first three steps are repeated until the new and old $\rho(\mathbf{G},t_2)$ are the same.
\end{enumerate} 

\subsection{Forces and dynamics}
{Once the self-consistency in charge density evolution is satisfied, post-processing including the calculation of total energy, Hellmann-Feynman forces, and the ionic trajectory are invoked. For instance, the forces acting on the ions can be calculated through 
	\begin{equation}
	\mathbf{F}_{\mathbf{R}_I} = \sum_{i\mathbf{k}} \braket{\psi_{i\mathbf{k}}|\nabla_{\mathbf{R}_I} \mathcal{H}|\psi_{i\mathbf{k}}},
	\end{equation}
	where $\mathbf{R}_I$ and $\mathbf{F}_{\mathbf{R}_I}$ are the position and force of $I$th ion.}

{With $\mathbf{R}_I$ and $\mathbf{F}_{\mathbf{R}_I}$, we utilize the Ehrenfest theorem for evolving ions according to the equation of motion \begin{equation}
	M_I \frac{d^2\mathbf{R}_I}{dt^2} = \mathbf{F}_{\mathbf{R}_I},
	\end{equation}
	where $M_I$ is the mass of $I$th ion. The velocity $v_I(t) = d\mathbf{R}_I/dt$ and the temperature $T(t) = \sum_I^{N_I} M_I v^2_I(t)/2N_I$ are also calculated, where $N_I$ is the total number of ions.}

{Besides the conventional NVE ensemble, additional thermostats, such as Nos\'{e}-Hoover~\cite{Nose1984,Hoover1985} and Berendsen~\cite{Berendsen1984} is considered to simulate different environmental conditions. In the damped MD simulations, we utilize a simple velocity-rescaling thermostat. The ionic velocities are rescaled at each time step as
	\begin{equation}
	\mathbf{v}_I'(t) = \mathbf{v}_I(t)\sqrt{T'(t)/T(t)},
	\end{equation}
	where $\mathbf{v}_I'(t)$ and $T'(t) = T(t) -\Delta T$ are the rescaled velocity and temperature, respectively. The decreasing rate $\Delta T = 0.01$~eV atom$^{-1}$ ps$^{-1}$ is used in the simulations.}

{Combining the TDKS equation and the Ehrenfest theorem, the many-body electron-electron interaction and the ionic movement under the excited-state TDKS wavefunction evolution are described in an \textit{ab initio} way. We expect that the electron-electron interactions at the adiabatic XC level and electron-phonon scatterings within the mean-field average trajectory are present in these simulations. The excess electronic energy could dissipate into available phonon modes via electron-phonon coupling or to low-energy electrons via electron-electron scattering, resulting in carrier thermalization and cooling effect.}

\subsection{Projector Augmented Wave Method}
To expand the TDKS orbital, the adiabatic basis $\ket{\phi_{i \mathbf{k}}(\mathbf{G},t_1)}$ should be orthonormal. Eigenstates in all electron method and norm-conserving pseudopotential satisfy the orthonormal requirement naturally. However, using projector augmented-waves method (PAW)~\cite{Blochl1994} bring additional core functions, which lead to nonorthogonal eigenstates. The formalism of DFT quantities such as total energy and forces using PAW method are described in Bl\"{o}chl's original paper~\cite{Blochl1994}. Here, we only list the related changes in TDDFT evolution.

Using PAW method, all the coefficients $C'_{\mathbf{k}}(t)$ are related with the pseudo-eigenstates $\ket{\tilde{\phi}_{i \mathbf{k}}(\mathbf{G},t_1)}$
\begin{equation}
\ket{\phi_{i\mathbf{k}}(\mathbf{G},t_1)} = \hat{T} \ket{\tilde{\phi}_{i\mathbf{k}}(\mathbf{G},t_1)}
\end{equation}
where $\hat{T}$ is the transform operator. The only change caused by PAW method is Eq~(\ref{innerProd}). Using PAW, eq~(\ref{innerProd}) becomes
\begin{equation}
\begin{split}
a_{il,\mathbf{k}}(t_1,t_2) &= \braket{\tilde\phi_{i\mathbf{k}}(t_1)|\hat{T}^{\dagger}\hat{T}|\tilde\phi_{l\mathbf{k}}(t_2)} \\
=&\braket{\tilde\phi_{i\mathbf{k}}(t_1)|\hat{S}|\tilde\phi_{l\mathbf{k}}(t_2)}
\end{split}
\end{equation}
where 
\begin{equation}
\hat{S} = \left(\begin{array}{cccc}
s_{11,\mathbf{k}}(t) &s_{12,\mathbf{k}}(t) & \cdots &s_{1N_b,\mathbf{k}}(t) \\
s_{21,\mathbf{k}}(t) &s_{12,\mathbf{k}}(t) & \cdots &s_{1N_b,\mathbf{k}}(t) \\
\vdots &\vdots & \cdots &\vdots \\
s_{N_b1,\mathbf{k}}(t) &s_{N_b2,\mathbf{k}}(t) & \cdots &s_{N_bN_b,\mathbf{k}}(t) \\
\end{array}\right),
\end{equation}
$s_{ij,\mathbf{k}} = \braket{\beta_{i\mathbf{k}}(\mathbf{G})|\phi_{j\mathbf{k}}(\mathbf{G})},$
and $\beta_{i\mathbf{k}}(\mathbf{G})$ is the Kleinman-Bylander projectors~\cite{Giannozzi2009}. 



\subsection{Time Dependent Band Unfolding}
The CDW phase of TiSe$_2$ is a $2\times2$ cell of the normal phase. The energy bands are folded from $1\times1$ Brillouin zone (BZ) to the $2\times2$ BZ. In contrast,  ARPES measurements still span over the $1\times1$ BZ. To bridge the gap between DFT bands and measured ARPES spectra, the band unfolding technique is used to calculate the effective band structure (EBS) of the supercell (SC). Expanding the adiabatic basis $\ket{\phi_{i,\mathbf{k}}(t)}$ of $2\times2$ SC in the adiabatic basis $\ket{\Phi_{I,\mathbf{K}}(t)}$ of primitive $1\times1$ cell (PC), we get 
\begin{equation}
\ket{\phi_{i,\mathbf{k}}(\mathbf{G},t)}=\sum_{I,\mathbf{K}}a(I,\mathbf{K};i,\mathbf{k};t)\ket{\Phi_{I,\mathbf{K}}\mathbf{G},t)},
\end{equation}
where $\mathbf{K}=\mathbf{k}+\mathbf{B}$ and $\mathbf{B}$ is the reciprocal basis vector of SC. 
The spectral function is the EBS along the $\mathbf{K}$ path in PCBZ~\cite{Popescu2012,Medeiros2014}:
\begin{equation}
A(\mathbf{K}, E, t)=\sum_i P(\mathbf{K};\mathbf{k},i;t)\delta(E-\epsilon_{i,\mathbf{k}}(t)),
\end{equation}
where $E$ is the energy and 
\begin{equation}
\label{probability}
P(\mathbf{K};\mathbf{k},i)=\sum_{I}a^*(\mathbf{K},I;\mathbf{k},i;t)a(\mathbf{K},I;\mathbf{k},i;t) =\sum_{\mathbf{G}}\left|\phi_{i,\mathbf{k}}(\mathbf{G}+\mathbf{K}-\mathbf{k}, t)\right|^2,
\end{equation}
We can introduce an extra weight function $w(i,\mathbf{k})$ 
\begin{equation}
\label{spacial_band_unfolding}
A(\mathbf{K},E,t)=\sum_i P(\mathbf{K};\mathbf{k},i;t) w_{i,\mathbf{k}}(t) \delta(E-\epsilon_{i,\mathbf{k}}(t)).
\end{equation} 
The choice of $w_{i,\mathbf{k}}$ is arbitrary~\cite{Lian2017SiUnfold}. Here, we use the population of the adiabatic states as the 
\begin{equation}
w_{i,\mathbf{k}}(t) = q_{i,\mathbf{k}}(t)
\end{equation}
to reproduce the intensity in ARPES spectra.  

\section{Fluence Dependence of Photocarrier Density}
\begin{figure}
	\centering
	\includegraphics[width=0.7\linewidth]{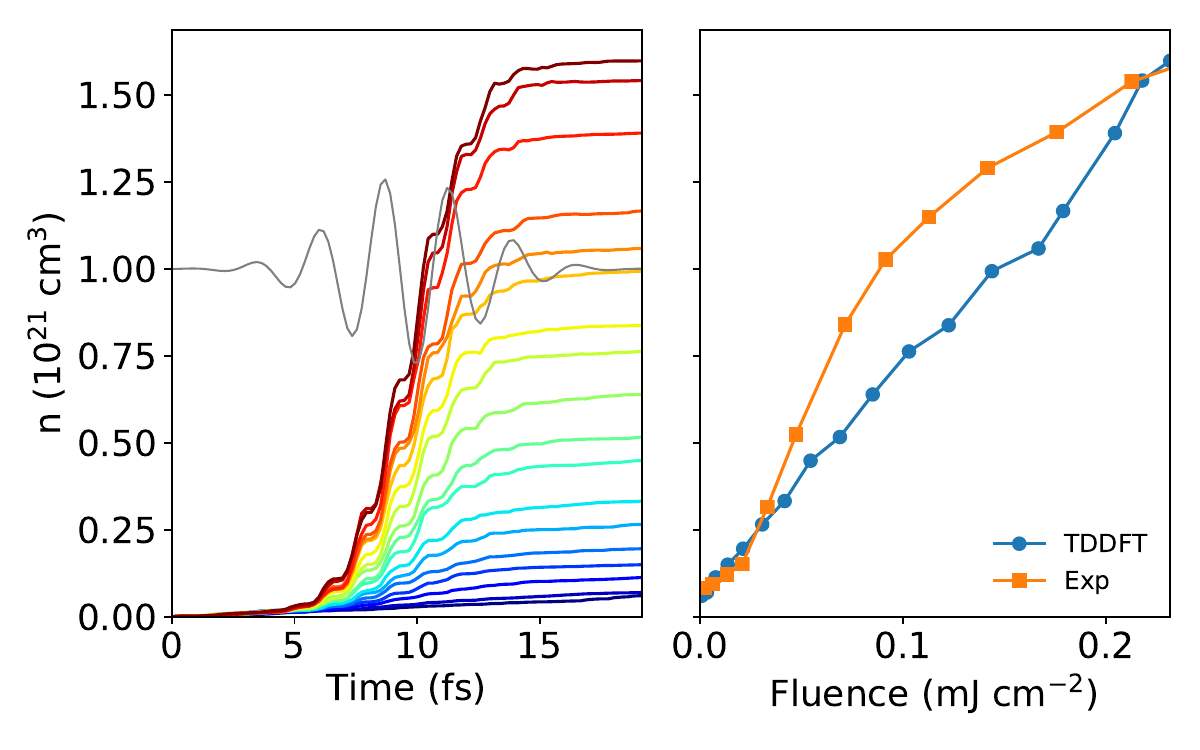}
	\caption{Photo-induced carrier density as (a) a function of time (b) a function of laser fluence. The grey line denotes the time-dependence of the laser pulse. The experimental data are reproduced from Ref.~\cite{Porer2014}. 
	}
	\label{fig:scanIntenCarrierwoExp}
\end{figure}

\begin{figure}
	\centering
	\includegraphics[width=0.7\linewidth]{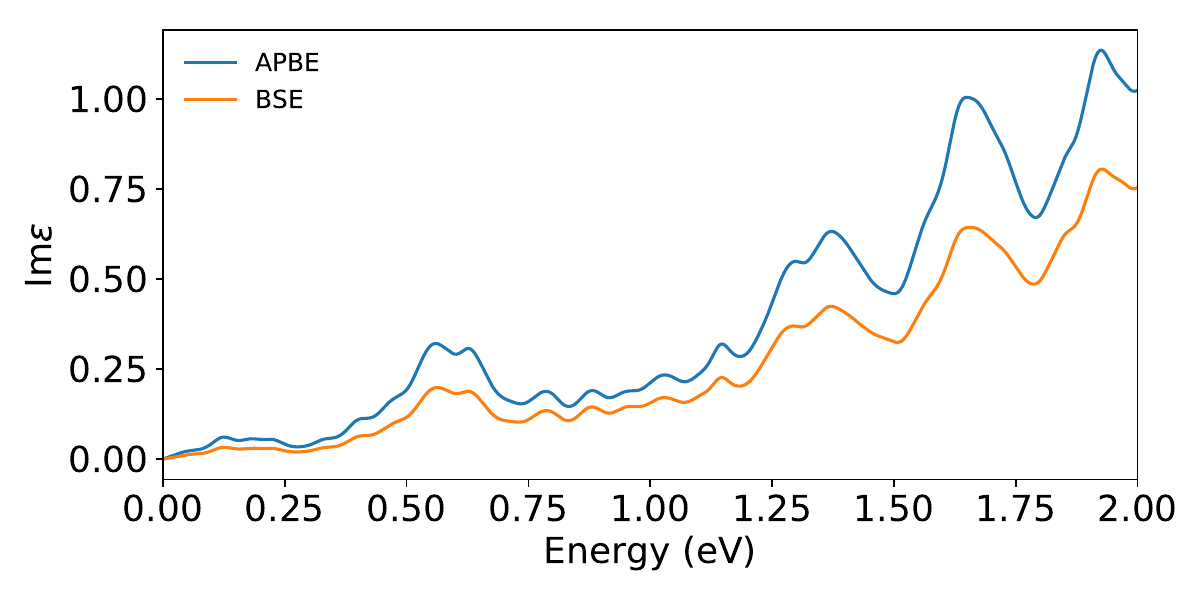}
	\caption{The imaginary part of the dielectric function of TiSe$_2$ at the momentum $\mathbf{q}=\mathbf{w}$ calculated from linear-response TDDFT with adiabatic PBE (APBE) and BSE kernels.}
	\label{fig:absComp}
\end{figure}

{We investigate the photo-carrier density as a function of time and laser fluence $n(t,I)$. Here, the carrier density is calculated as $n(t) =  \frac{1}{2}\sum_{i,\mathbf{k}}|q_{i\mathbf{k}}(t)-q_{i\mathbf{k}}(t=0)|$. We note that $n(t)$ accurately describes the number of excited carriers after the laser field ends $A(t)=0$. Otherwise, a gauge independent projection on $\psi_{k'}(t)$ with $k'=k-A(t)/c$ instead of $k' = k$ can be used~\cite{Otobe2008, Otobe2009, Otobe2016b, Yabana2012, Shinohara2010, Shinohara2010a, Shinohara2012, Sato2015, Sato2015a}. These two projections are identical when the laser field ends $A(t)=0$.}

As shown in Fig.~\ref{fig:scanIntenCarrierwoExp}, the $n(t=20~\mathrm{fs},I)$ are approximately proportional to the laser fluences. After comparing with the experimental measurements~\cite{Porer2014}, we find the $n(I)$ relations are consistent at both the low fluence area $I<0.04$~mJ cm$^{-2}$ and the high fluence area $I>0.22$~mJ cm$^{-2}$, while a clear derivation from linearity is observed in experiments. Exciton correlation drives the superlinear feature but is underestimated in DFT/TDDFT calculations (Fig.~\ref{fig:absComp}). Nevertheless, since the difference is minimized when $I>0.22$~mJ cm$^{-2}$, we focus on the experimental phenomena in high laser fluence region, to reproduce the similar excitation states for direct comparisons. 

{Consistent with experimental observations~\cite{Mohr-Vorobeva2011}, we note that the PLD dynamics are not sensitive to the photon energies. This is because that the band gap of CDW 1T-TiSe$_2$ (0.18~eV) is smaller than the photon energy in most commonly used laser sources ($\sim$1~eV). Besides, the sub-picosecond laser pulses utilized in the experiments and our simulations would bring up significant broadening in photon energy as well as multi-photon absorption processes.}

\end{document}